ORIGINAL ARTICLE

# Single conjugate adaptive optics for the ELT instrument METIS

**Stefan Hippler, et al.** ***[full author details at the end of the article]***

Received: 15 November 2017 / Accepted: 17 October 2018 / Published online: 20 November 2018

**Abstract**

The European Extremely Large Telescope (ELT) is a 39 m large, ground-based optical and near- to mid-infrared telescope under construction in the Chilean Atacama desert. Operation is planned to start around the middle of the next decade. All first light instruments will come with wavefront sensing devices that allow control of the ELT's intrinsic M4 and M5 wavefront correction units, thus building an adaptive optics (AO) system. To take advantage of the ELT's optical performance, full diffraction-limited operation is required and only a high performance AO system can deliver this. Further technically challenging requirements for the AO come from the exoplanet research field, where the task to resolve the very small angular separations between host star and planet, has also to take into account the high-contrast ratio between the two objects. We present in detail the results of our simulations and their impact on high-contrast imaging in order to find the optimal wavefront sensing device for the METIS instrument. METIS is the mid-infrared imager and spectrograph for the ELT with specialised high-contrast, coronagraphic imaging capabilities, whose performance strongly depends on the AO residual wavefront errors. We examined the sky and target sample coverage of a generic wavefront sensor in two spectral regimes, visible and near-infrared, to pre-select the spectral range for the more detailed wavefront sensor type analysis. We find that the near-infrared regime is the most suitable for METIS. We then analysed the performance of Shack-Hartmann and pyramid wavefront sensors under realistic conditions at the ELT, did a balancing with our scientific requirements, and concluded that a pyramid wavefront sensor is the best choice for METIS. For this choice we additionally examined the impact of non-common path aberrations, of vibrations, and the long-term stability of the SCAO system including high-contrast imaging performance.

**Keywords** Single conjugate adaptive optics · SCAO · ELT · Pyramid wavefront sensor · Shack-Hartmann wavefront sensor · Fragmented pupil · Low wind effect · Non-common path aberrations · High-contrast imaging

## 1 Introduction

The mid-infrared ELT imager and spectrograph METIS is one of three first instruments on the European Extremely Large Telescope (ELT) [1]. The ELT is currently under construction with an estimated completion date around 2025. The other 2 first light instruments are MICADO [2], a near-infrared, 0.8–2.4 $\mu$m, imager and spectrograph, and HARMONI [3], an integral field spectrograph sensitive in the 0.47–2.45 $\mu$m regime. All 3 first ELT instruments come with adaptive optics tuned to the scientific requirements and goals of each instrument. For MICADO this translates for example in the design of the multi-conjugate adaptive optics system MAORY [4], while for HARMONI a laser tomography adaptive optics (LTAO) system is foreseen. For an overview of AO in astronomy see review article by Davies & Kasper [5]. An overview of the currently planned AO systems for the next generation of extremely large telescopes can be found in [6].

METIS covers the mid-infrared/thermal spectral range between 2.9 –19 microns. Diffraction limited imaging, coronagraphy, medium resolution ($R \sim 10^2 - 10^3$) slit spectroscopy over the full spectral range (starting at 3 $\mu$m) and high resolution ($R \sim 10^5$) integral field spectroscopy in the lower spectral range (2.9 – 5.3 $\mu$m) make METIS a versatile instrument [7]. The compact imaging field of view of $\sim 10'' \times 10''$ together with a much larger isoplanatic angle of about 20$''$ for the shortest science wavelength and median atmospheric conditions (Table 4), clearly indicated the use of a single conjugate adaptive optics (SCAO) system to achieve diffraction limited performance [8].

Starting with the choice of the wavefront sensor's spectral range in Section 3, we describe the simulation tool we used to estimate the SCAO performance in Section 4. In Section 5 we describe in detail the parameters used for the simulations and in Section 6 we present the simulation results. A one hour simulation of a representative METIS observation of the exoplanetary system 51 Eri is presented in Section 7. We show the obtained coronagraphic point spread functions and a corresponding contrast curve. Section 8 contains our conclusions and outlines the next steps until the preliminary design review of the METIS instrument, which is foreseen to take place in spring 2019.

## 2 Requirements

The scientific requirements of the METIS instrument, which are relevant for the design of the SCAO system are:

– Minimum Strehl ratio (R-MET-111): METIS and its associated natural guide star SCAO system shall deliver at least 93% Strehl (goal: 95%) at 10 $\mu$m, and at least 60% (goal: 80%) Strehl at 3.7 $\mu$m. These numbers are based on nominal ELT optics, a median V-band seeing of 0.65”, a zenith angle of 30 degree, and a natural guide star with $m_K = 10$ mag. This performance shall be provided continuously over at least 15 min under nominal telescope operating conditions. This and all other numbers are valid for the science focal plane, i.e. they include the correction of static and non-common path aberrations. The balancing with components in the beam that quasi-deliberately worsen these numbers is still under discussion.

- Off-zenith observations (R-MET-119): METIS and its natural guide star SCAO shall be able to provide AO correction up to 60 degree zenith angle with less than 40% degradation of the Strehl ratio (with respect to zenith) for a bright star ($m_K \approx 8$ mag) under median seeing conditions.
- High-contrast imaging (HCI): in order to facilitate high-contrast observations, METIS and its natural guide star SCAO shall guarantee a residual image motion in the coronagraphic focal plane of less than 5 mas rms (goal: 2 mas rms) under the conditions outlined in R-MET-111. Note that image motion is wavelength independent.

A compact overview of the METIS science cases can be found in [9]. The METIS requirements in this chapter are part of the METIS technical specification document [10, 11].

## 3 Spectral range for wavefront sensing

An important decision for the design of the METIS SCAO system was the choice of wavefront sensing wavelength. Various factors play a role in this context. Besides the underlying detector technology, one key factor is the sky and sample coverage. Sky coverage is a statistical number that defines the probability to find over the whole sky a sufficiently bright reference source for the wavefront sensor (WFS). In contrast, the sample coverage defines the probability to use targets from a given scientific sample as wavefront sensor reference source, or to find one sufficiently close. The choice depends on the instrument philosophy: is it a general purpose facility type instrument or is it targeting specific science cases.

To estimate the sky coverage we can look at the flux emitted from main sequence stars at a distance of 100 pc as a function of stellar mass. Here we consider 100 pc as the largest distance suitable for one of the major science goals of METIS, which is direct imaging and characterisation of exoplanets. The lower the exoplanet host stellar mass, the lower the temperature, the higher the flux in the infrared compared to the shorter wavelength regime. The opposite is true for high mass exoplanet host stars. In our detailed performance simulations (Section 6), we find that we achieve very high adaptive optics performance at detected near-infrared flux levels down to $\sim$100 photons per wavefront sensor integration time and sub-aperture size. For METIS, this corresponds to $8{\cdot}10^5$ photons/s/m$^2$ in K-band, corresponding to stars with stellar mass higher than about 0.7 solar masses. For low mass stars, Fig. 1 shows that there is a “flux” advantage going to the near-infrared regime.

As shown and discussed in [8, 14], there is another advantage using the near-infrared spectral range for wavefront sensing in combination with a pyramid WFS. The reason for dividing the spectral range into a visible 0.6–1$\mu$m regime and a near-infrared 1–2.5$\mu$m one is driven by the available detectors in the respective regime, e.g. CCD detectors or HgCdTe focal plane arrays. It has already been demonstrated

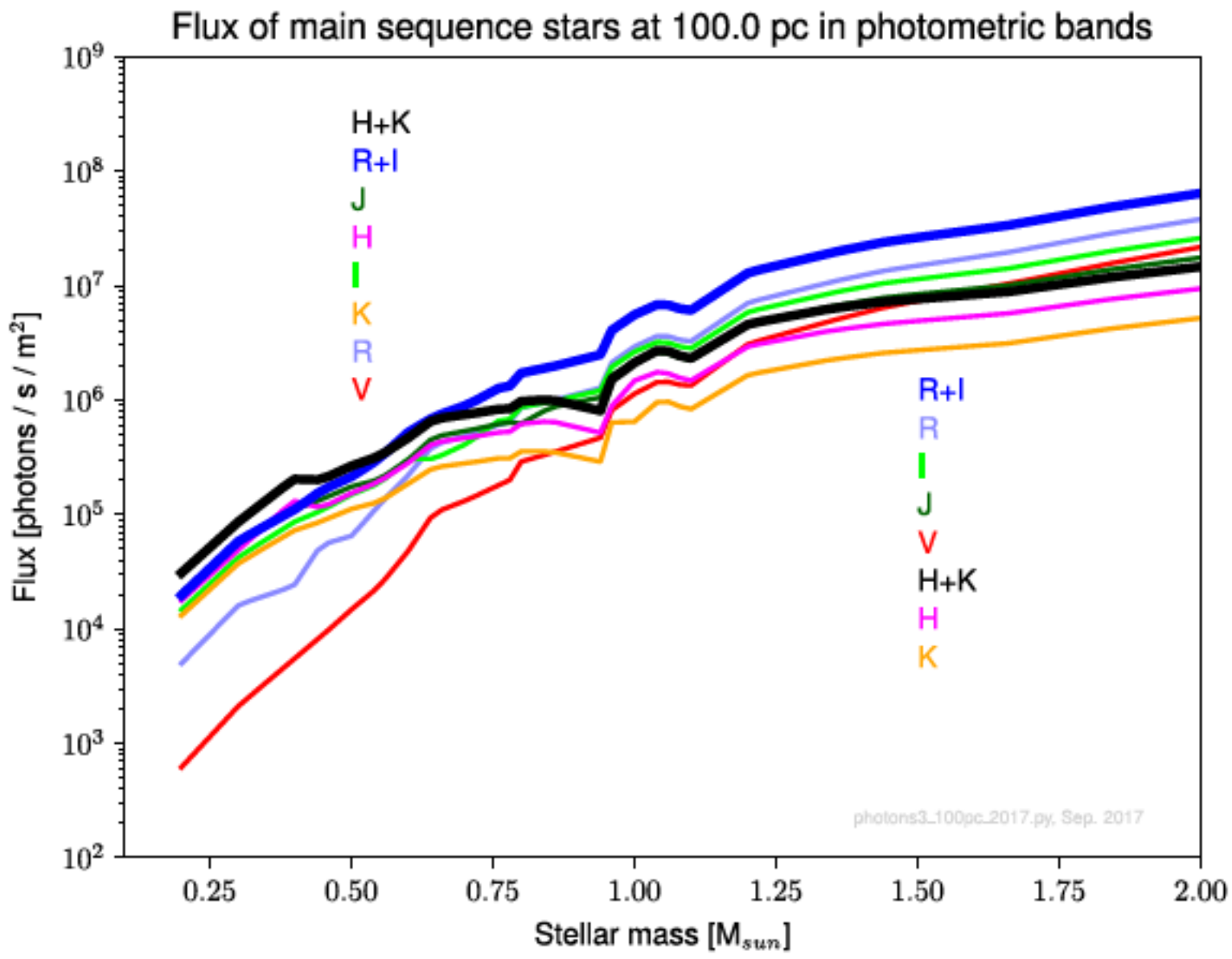

**Fig. 1** Photon flux of main sequence stars at a distance of 10 parsec as a function of stellar mass and spectral band. Spectral band name, central wavelength $\lambda_c$ and bandwidth $\Delta_\lambda$ are: **V**, $\lambda_c$=0.56 $\mu$m, $\Delta_\lambda$=0.09 $\mu$m; **R**, $\lambda_c$=0.7 $\mu$m, $\Delta_\lambda$=0.22 $\mu$m; **I**, $\lambda_c$=0.9 $\mu$m, $\Delta_\lambda$=0.24 $\mu$m; **J**, $\lambda_c$=1.25 $\mu$m, $\Delta_\lambda$=0.3 $\mu$m; **H**, $\lambda_c$=1.65 $\mu$m, $\Delta_\lambda$=0.35 $\mu$m; **K**, $\lambda_c$=2.2 $\mu$m, $\Delta_\lambda$=0.4 $\mu$m. Stellar flux magnitudes taken from [12]. Conversion to photons per second and square meter according to [13]

that wavefront sensing in R-band works excellent for imaging in N-band (8–13 $\mu$m) [15, 16].

Chromatic correction errors, i.e. optical path differences due to the quite wide distance between the WFS spectral band and the actual METIS observation spectral band between 3–19$\mu$m are for typical seeing conditions about 40 nm [17].

Interestingly, the ratio of atmospheric coherence time to control loop latency time increases with longer wavelengths (latency is wavelength independent). With a fixed AO control frequency, this can also be seen as an advantage of a near-infrared WFS.

Furthermore, the transmissive optical elements of METIS must also be considered to determine the best possible spectral range for the WFS. As shown in Fig. 2, the key optical elements for the SCAO unit are the entrance window of the METIS cryostat and the internal dichroic beam-splitter between the SCAO unit and the METIS spectrograph and imager. The already very wide spectral range of METIS becomes even wider due to the adaptive optics working in the outside spectral range, at shorter wavelengths. While this approach can make the best possible use of the stellar photons, the optics must also allow this. Fortunately, this is possible as shown in Fig. 3 for the case of using a near-infrared WFS. A similar approach is used in the ESO VISIR instrument upgrade NEAR [20], where a dichroic beam-splitter reflects the visible spectrum to the VISIR AO system and transmits the N-band (8–13 $\mu$m) to the science channel.

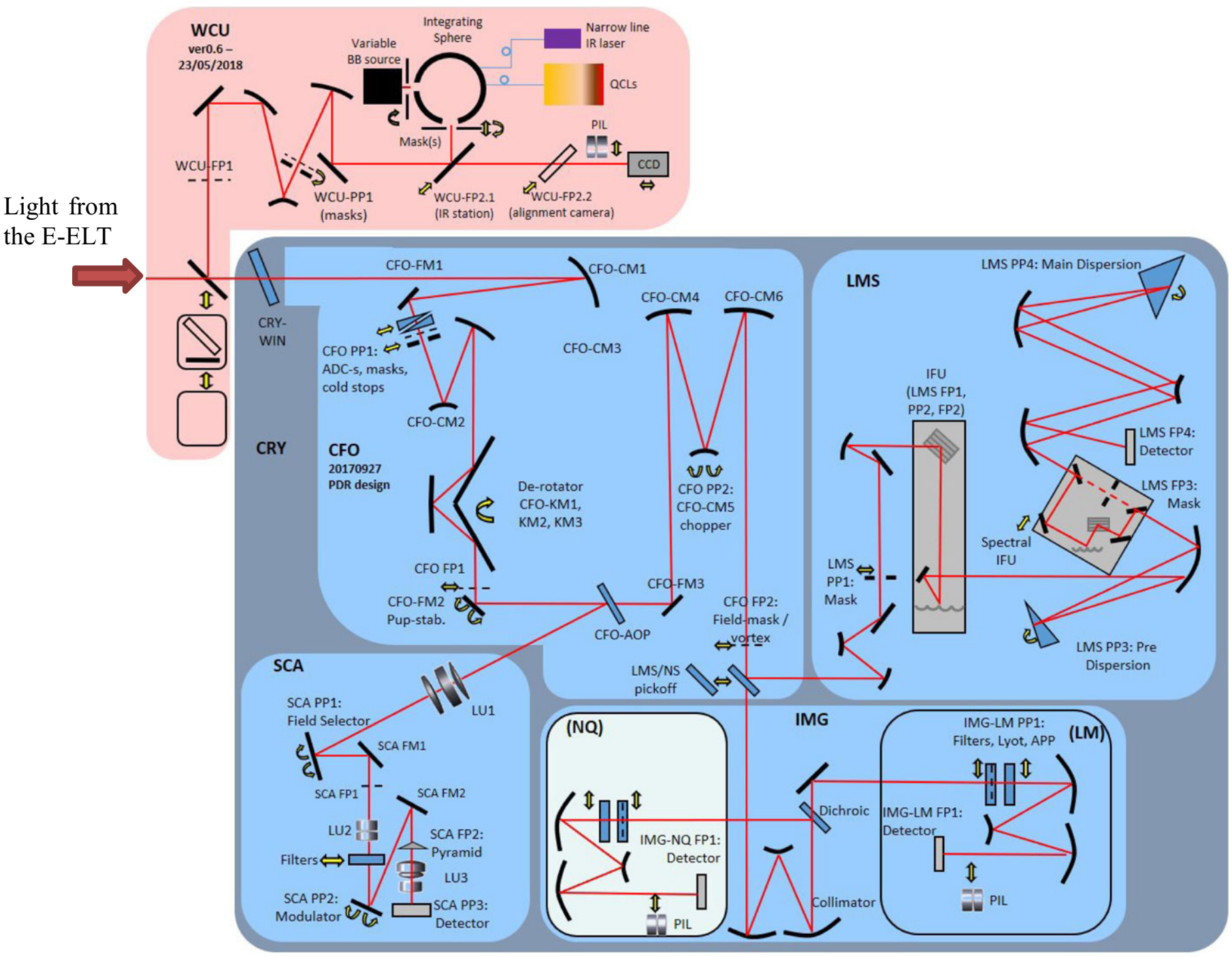

**Fig. 2** Optical overview of METIS taken from [18]. Light from the telescope enters the METIS cryostat from the left. Alternatively, light from a warm calibration unit (WCU) can be used. For this, a movable beam-splitter can be inserted in the optical beam. The first transmissive element with unused WCU is the entrance window of the METIS cryostat (CRY-WIN). All components of the upper left blue box, with the exception of an atmospheric dispersion compensator (ADC) are reflective and are located in the common beam path, the so-called common fore optics (CFO). A beam-splitter (CFO-AOP) reflects part of the star light into the SCAO (SCA) unit. The transmitted light is relayed via further reflective elements into the spectrograph (LMS, upper right box) or imager (IMG, lower right box). The components of the WCU (red colored box) as well as the spectrograph and imager are irrelevant for further consideration. More details can be found in [18, 19]

At the very end, all factors that have an impact on the AO performance have to be considered to come to a sound decision for the wavefront sensor's spectral band. We tried to narrow down this rather extensive task by looking at the signal to noise ratio (SNR) per sub-aperture of a generic wavefront sensor with 2 types of real-world detectors, an electron multiplying CCD (EMCCD [21]) and an electron avalanche photodiode near-infrared detector (SAPHIRA [22, 23]).

We analysed two target samples, one containing 232 late-type stars (spectral type M5, J$\leq$10 mag, DEC $\leq$ +20 deg, selected from the bright M-dwarf sample of Lépine & Gaidos [24]), the other containing 15126 main sequence stars within 100 parsec taken from the Hipparcos catalogue [25]. The selection criterion is to achieve a given minimum SNR per sub-aperture. For small SNR values, we find that with this criterion both samples can be equally well observed with a visible or near-infrared based wavefront sensor. For SNR values around 5 (see Fig. 4 for SNR=4), there is an advantage using a near-infrared wavefront sensor. For high SNR values, there is no obvious choice. The results for SNR values up to 18 are shown in Fig. 5.

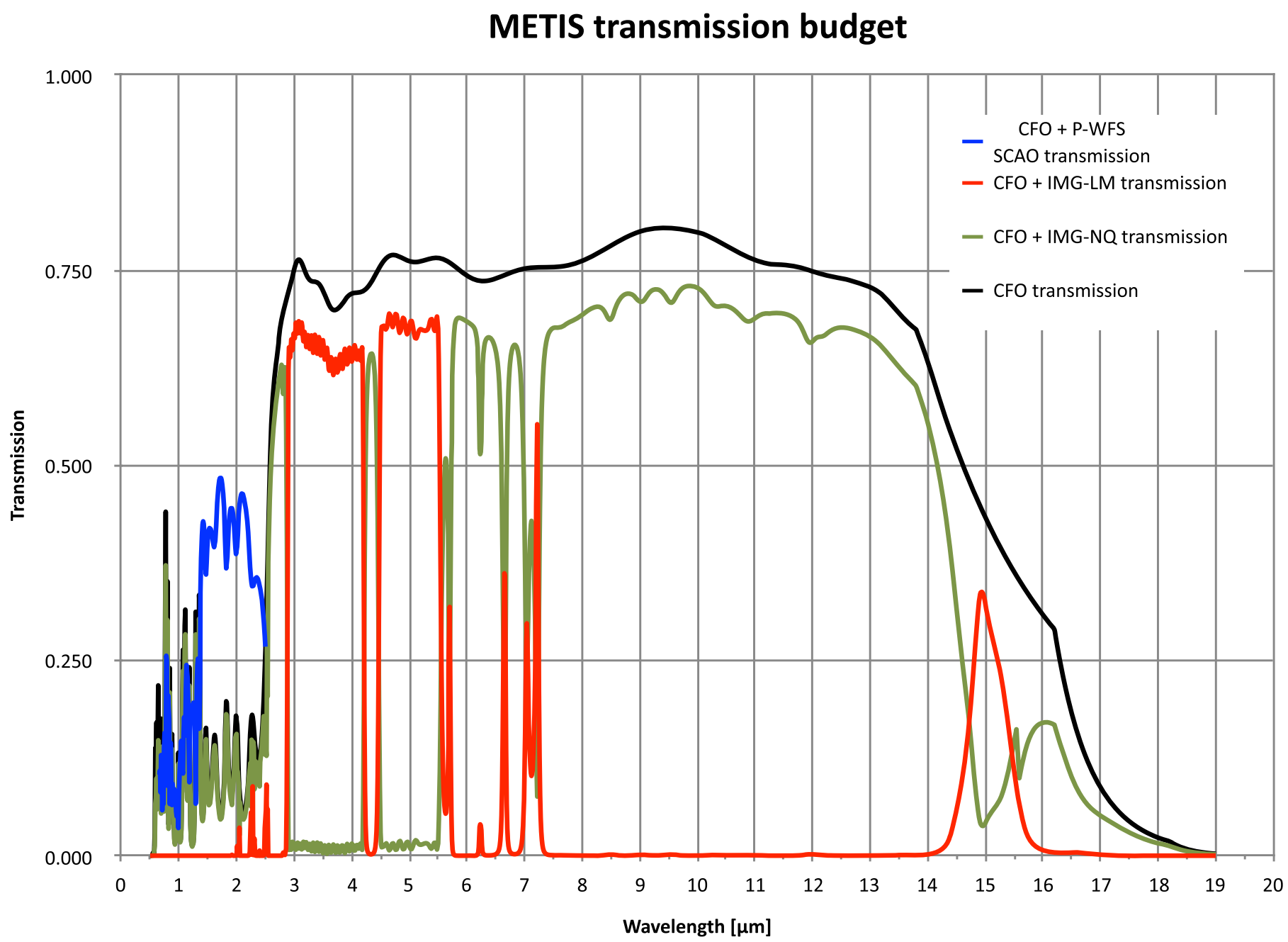

**Fig. 3** METIS transmission budget for a near-infrared SCAO unit. The black line shows the transmission of the common fore optics (CFO) including the cryostat entrance window. The blue line shows the CFO transmission and the SCAO transmission until just before the pyramid WFS (P-WFS) detector. The green and red lines show the transmission until just before the detectors of the LM-band ($\sim$ 3–5 $\mu$m) and NQ-band ($\sim$ 7–19 $\mu$m) imagers resp.

In view of the results shown in Figs. 1, 4, and 5, together with our results published in [8, 14], we conclude that the near-infrared wavelength regime is the preferential choice for the METIS SCAO wavefront sensor. We further preselected the precise near-infrared spectral range to be equal to the one used for the near-infrared wavefront sensors at the VLT interferometer [26, 27], to be specific 1.4$\mu$m – 2.4$\mu$m including H-band and K-band. Once this choice is confirmed, one can analyse the benefit of adding J-band (centered around 1.25 $\mu$m) to the wavefront sensor's spectral band. Although the optical components of the METIS instrument are not optimised for such short wavelengths, and the optical throughput in J-band is very low, there is a significant gain in sky and sample coverage for high mass reference stars. This gain is summarized in Table 1.

## 4 The adaptive optics simulation tool yao

After using the PAOLA [28] AO simulation tool during the METIS phase A study [29], we switched to yao [30], an open-source, general-purpose AO simulation tool written by François Rigaut in the interpreted programming language yorick [31].

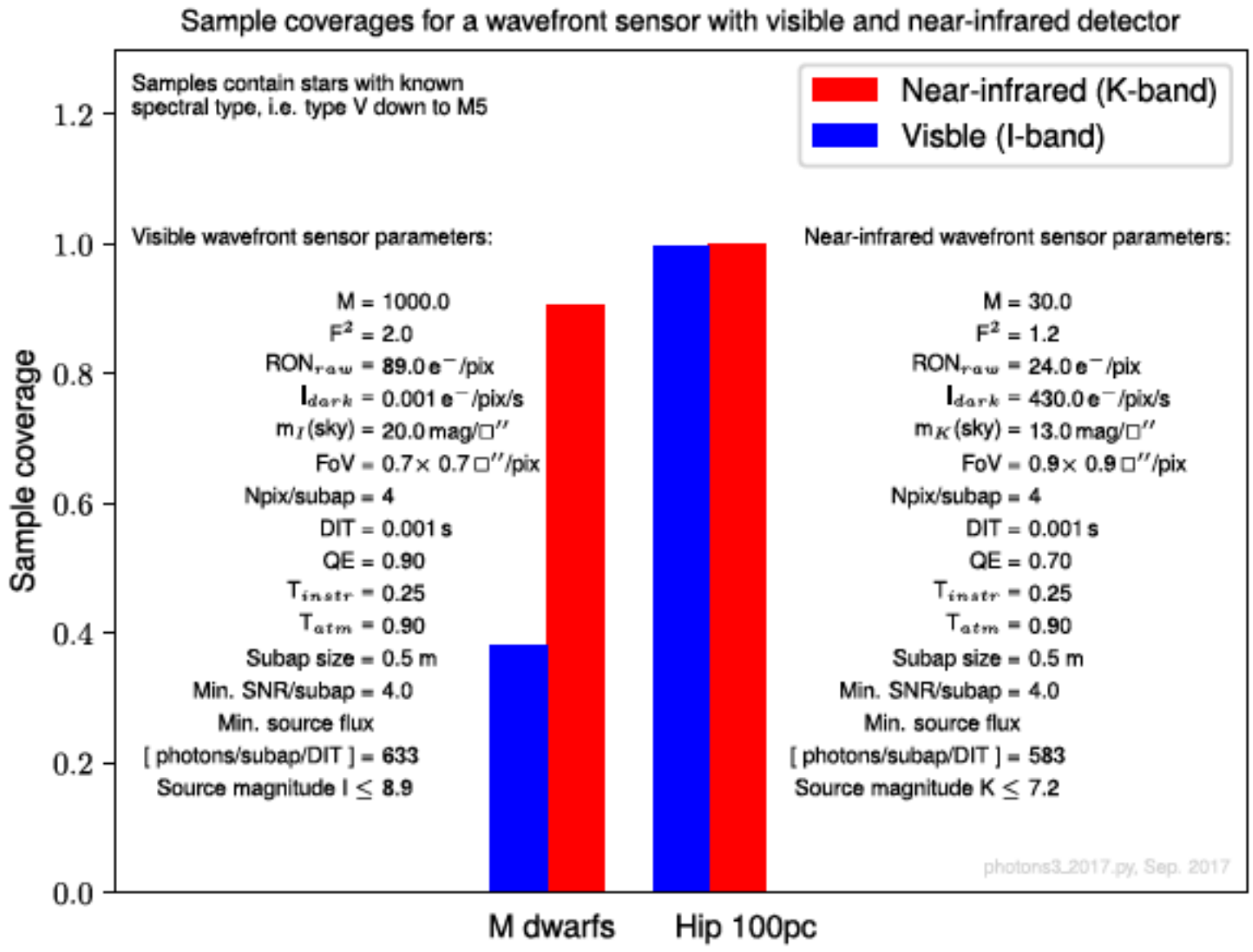

**Fig. 4** Exemplary wavefront sensor sample coverages with a near-infrared and visible detector and for a minimum signal to noise value per sub-aperture of 4. The number of pixels per sub-aperture Npix/subap, detector integration time $DIT$, detector quantum efficiency $QE$, instrument transmission $T_{instr}$, atmospheric transmission $T_{atm}$, and sub-aperture (subap) size are identical for both types of detectors, the visible one, with its spectral range in I-band and the near-infrared one, with its spectral range in K-band. Different parameters are the detector amplification gain $M$, the amplification noise factor $F^2$, the detector raw read-out noise $RON_{raw}$, the detector dark current $I_{dark}$, and the pixel field of view FoV. Also different are the sky background levels for the respective wavefront sensor spectral band, $m_I$ and $m_K$

In this section, we report on the important upgrades and modifications made to yao for our purposes. For many, but not all simulations, the latest available version, yao 5.10.2 running in yorick 2.2.04x, was used. For some simulations the original code

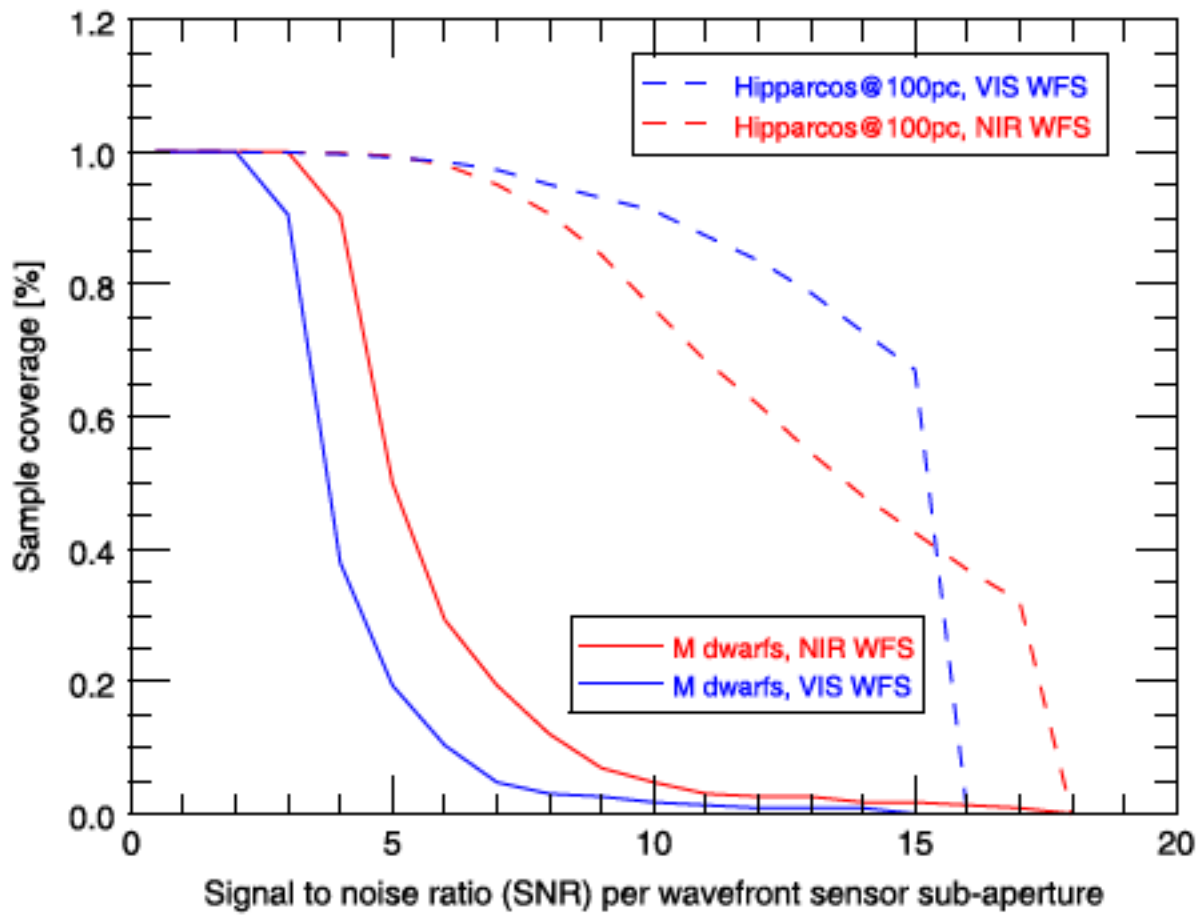

**Fig. 5** Wavefront sensor (WFS) sample coverages with a near-infrared (NIR) and visible (VIS) detector. The detector integration time is 1 ms. For other details see Fig. 4

**Table 1** Gain in observable objects for a wavefront sensor spectral range including J-band, i.e. spectral range 1.1–2.4$\mu$m (JHK-bands) vs. spectral range 1.4–2.4$\mu$m (HK-bands), for different classes of reference stars

| Reference star spectral type | J-H mag | H-K mag | Gain in number of photons | Gain in number of observable objects |
|---|---|---|---|---|
| A0V | 0 | 0 | 2.4 | 3.7 |
| G2V | 0.27 | 0.07 | 2.1 | 3.0 |
| M0V | 0.65 | 0.18 | 1.7 | 2.3 |
| M5V | 0.61 | 0.29 | 1.7 | 2.3 |

For simplicity we assumed that the quantum efficiency of the detector and the instrumental throughput is flat over the 1.1–2.4$\mu$m spectral range, and a constant density of the reference stars in space around the Sun for each spectral class was modified to incorporate for example non-common path aberrations or influence functions of the ELT's deformable mirror M4.

Yao can be fully controlled with one parameter file. Within this file, the atmosphere, the wavefront sensor(s), the deformable mirror(s), the reconstruction method, the AO loop parameters, the wavefront sensor wavelength, the science target wavelength(s), and the wavefront sensor's guide star brightness are specified. In our simulations, we start with two slightly different yao parameter files, one that configures a Shack-Hartmann wavefront sensor (SHS, [32]) and one for a Pyramid wavefront sensor (PYR, [33]). In a second step, we use scripts that vary parameters like the seeing, the guide star magnitude, the zenith angle of the guide star, the AO loop gain, the AO loop frequency, the sub-aperture size, a regularisation parameter, that controls the inversion of the interaction matrix as explained in Section 4.1, and the detector pixel threshold below which pixel values are disregarded in the wavefront slope computation. This allows to find the optimal configuration in terms of residual wavefront error or Strehl ratio (SR) as, for example, a function of guide star magnitude.

### 4.1 Wavefront reconstruction

Wavefront reconstruction with yao allows to use 3 methods to build the adaptive optics control matrix. For all simulations discussed below we used the minimum mean square error (MMSE) reconstructor. For a Shack-Hartmann or Pyramid wavefront sensor, estimating the wavefront error $\phi$ from wavefront slopes $s$ can be formulated with the linear equation

$$s = H\phi + n, \tag{1}$$

where $n$ is the noise vector and $H$ the interaction matrix. $H$ is generated in yao using either internally generated deformable mirror influence functions or user defined influence functions. In our simulations we used both yao internal influence functions generated for a stack-array deformable mirror, and modelled ELT M4 influence functions provided by ESO. In both cases, the calibration scheme follows a "zonal" approach, i.e. during initialisation, yao applies each actuator's influence function and measures the response of the wavefront sensor. In this way each "actuator" column of the interaction matrix $H$ is filled with a slope vector of size twice the number of wavefront sensor sub-apertures, i.e. x- and y-slopes for all valid sub-apertures. Using the MMSE method [34] in yao to create the AO control matrix means solving (1), i.e. find a matrix $R$ that gives a good estimate of

$$\phi = Rs \tag{2}$$

with the well known result (see for example [35])

$$R_{mmse} = (H^T H + aC)^{-1} H^T, \tag{3}$$

where $C$ is a regularisation matrix, $a$ a regularisation parameter, and the superscript symbols T and -1 stand for the transpose and inverse of a matrix. For a = 0, (3) reduces to the well known least-squares wavefront estimator, $R_{LS} = (H^T H)^{-1} H^T$. The regularisation matrix $C$ is either user provided or the identity matrix. Optionally, in case of a stack-array piezoelectric deformable mirror $C$ can be created by convolving a laplacian operator by itself (yao parameter dm.regtype = "laplacian"). See Fig. 6 for an example of a stack array actuator map and its corresponding Laplacian regularisation matrix. In the latter case, the regularisation "matrix has similar statistics to the inverse covariance matrix for Kolmogorov turbulence and penalises local waffle in the deformable mirror" [36].

In closed-loop, the control matrix together with the WFS measurements is used to compute the control vector, e.g. the control voltages for the DM. Both, the actual

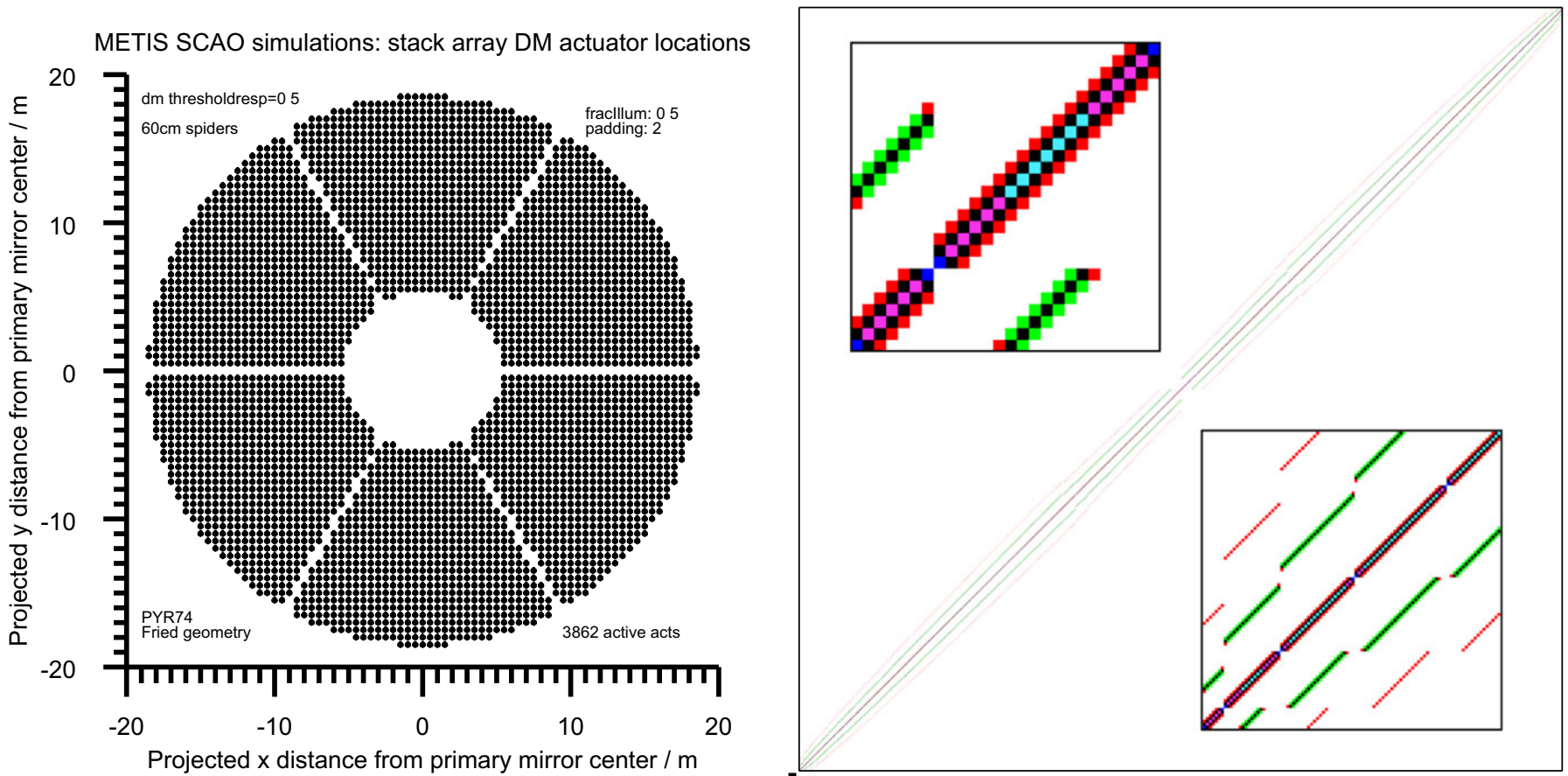

**a** Yao generated actuator map after calibration with a piezo stack array DM and the ELT circular ring shaped pupil mask with central obscuration and 60 cm wide spiders (section 5.4). A pyramid wavefront sensor (PYR74) with an actuator response threshold of 0.5 was used (see sections 5.1 and 5.7 for details). The 3862 active actuators are numbered from left to right starting with the bottom row.

**b** Regularisation matrix of size 3862 x 3862 used with stack array mirrors. Lower left corner 26 x 26 matrix (first 26 actuators) extract is shown in the upper left inset. Lower left corner 100 x 100 (first 100 actuators) matrix extract is shown in the lower right inset. Different colours correspond to matrix values in the range -0.5 to 1.25, white colour to value 0.

**Fig. 6** Actuator location map (**a**) and regularisation matrix (**b**) used in yao with a stack array deformable mirror. The actuator pitch corresponds to 0.5 m on the 37 m ELT pupil

WFS measurement as well as the control vector computation take time. The consequence of this is that the wavefront correction takes place with a time delay or latency. Latencies have an impact on loop stability and AO performance (see for example [37]). Section 5.8 outlines how yao handles this.

The combination of regularization parameter $a$, the loop gain $g$, and a detector pixel threshold value $t$ are in this study always optimized for all parameter combinations under study. Optimisation in our case is restricted to selecting the best possible value from a pre-defined set. The other available reconstruction methods in yao are singular value decomposition (SVD) and a sparse matrix version of MMSE. We further implemented a cumulative reconstructor for a Shack-Hartmann sensor, CURE-D [38] in yao, by far the fastest method for computation of the control voltages but with slightly reduced AO performance results compared with the standard yao reconstructors. A similar reconstructor for pyramid wavefront sensors exists [39] but was not implemented in our yao simulations.

## 5 Simulation parameters

### 5.1 Simulated wavefront sensors and sub-aperture sizes

Three wavefront sensor types were investigated in our simulations in order to support a decision between them. Details of their configurations (SHS60, SHS74, PYR74) are listed in Tables 2 & 3 as well as in the text below. The numbers in these acronyms stand for the linear number of sub-apertures used for wavefront sensing. They correspond to linear sub-aperture sizes of 0.62 m for the number 60 and 0.5 m for the number 74. The number of sub-apertures or actuators is often given in the text as a linear number over the telescope diameter/pupil. For the simulation, the corresponding square numbers are used according to a two-dimensional pupil. The specification of 74 linear sub-apertures, for example means that a maximum of $74 \times 74$ sub-apertures are used.

In this work only the 2 sub-aperture sizes mentioned above (see also Table 2) are investigated. Usually, a sufficiently good choice is to adjust the sub-aperture size according to the Fried parameter $r_0$. For METIS and median seeing conditions (see

**Table 2** Simulated wavefront sensors

| Simulation name | Sensor type | Spatial sampling [m] | Temporal sampling [Hz] | Detector |
|---|---|---|---|---|
| SHS60 | Shack-Hartmann | 0.62 | up to 1000 | SAPHIRA |
| SHS74 | Shack-Hartmann | 0.5 | up to 1000 | to be defined |
| PYR74 | Pyramid | 0.5 | up to 1000 | SAPHIRA |

For a Shack-Hartmann wavefront sensor with 74 sub-apertures across the ELT pupil and a minimum number of 4 detector pixels across each sub-aperture, the existing SAPHIRA detectors do not provide enough pixels

**Table 3** Yao parameters used for the SHS60, SHS74, and PYR74 simulations using a 37 m circular ring masked pupil, ELT segments, no spiders for the SHS systems and 60 cm wide spiders for PYR74

| Yao parameter | SHS60 | SHS74 | PYR74 |
|---|---|---|---|
| Linear size of simulation grid in pixels | 300 | 370 | 370 |
| Pupil mask used (Section 5.4) | ELT pupil without spiders | same as SHS60 | ELT pupil with 60 cm wide spiders |
| Linear number of pixels per sub-aperture on simulation grid | 5 | 5 | 5 |
| Number of sub-apertures across the pupil | 60 | 74 | 74 |
| Padding on each side of the simulation grid in # of sub-apertures | — | — | 2 |
| Total number of sub-apertures used | 2472 | 3816 | 4160 |
| Minimum sub-aperture illumination in percent | 90 | 90 | 50 |
| Linear number of detector pixels to find spot location | 4 | 4 | — |
| Detector pixel field of view in arcsec/pixel | 0.69 | 0.85 | — |
| Detector read-out noise in electrons/pixel | 1 | 1 | 1 |
| Detector dark current in electrons/pixel/s | 1000 | 1000 | 1000 |
| Total number of points along beam modulation circle | — | — | 12, 24 |
| Modulation amplitude from center of pyramid in arcsec | — | — | 0.05 |
| Field stop shape | round | round | round |
| Field stop size in arcsec | 2.76 | 3.4 | 1.8 |
| Modulator location | — | — | after field stop |
| Detector pixel threshold in electrons/pixel | 0–3 | 0–3 | 0–3 |
| Centroiding algorithm | CoG with yao default pixel thresholding | same as SHS60 | QC formula with yao default pixel thresholding |
| Actuator calibration amplitude in micrometer | 1.0 | 1.0 | 0.1, 0.25, 1.0 |
| Tip-tilt mirror calibration amplitude in mas | 200 | 200 | 20, 50, 200 |
| High-order DM optical conjugation | ground (0 m) | same as SHS60 | same as SHS60 |

**Table 3** (continued)

| Yao parameter | SHS60 | SHS74 | PYR74 |
|---|---|---|---|
| Tip-tilt mirror optical conjugation | ground (0 m) | same as SHS60 | same as SHS60 |

For the SHS systems, the centroiding algorithm uses the center of gravity (CoG) method with the default yao pixel thresholding method. The given pixel threshold is subtracted from all pixel values and resulting negative pixel values are set to zero. For PYR74, the centroiding algorithm uses the standard quad cell (QC) centroiding method with the default yao pixel thresholding method. Pixels with values below the threshold are set to the threshold value

Section 5.3), $r_0$ is larger than 1 m even for the shortest wavelength of 3 $\mu$m. The reasons that we have selected 2 sizes much lower as possibly necessary are:

a) a sub-aperture size of 0.5 m equals the average actuator spacing of the ELT deformable mirror and therefore matches the spatial characteristics of wavefront sensing and correction.
b) the 0.62 m sub-aperture size was chosen in order to be compliant with actually available and used near-infrared detectors [23], i.e. a Shack-Hartmann sensor with 60 sub-apertures and 4 linear pixel per sub-aperture just fits the size of the SAPHIRA device with 320 × 256 pixel [22].
c) the wavefront sensor itself operates in diffraction or nearly-diffraction limited mode.

A more detailed trade-off study to balance out the sub-aperture size with the WFS sensitivity and AO performance is planned for 2019. The main parameters of the investigated wavefront sensors in this paper are listed in Table 2.

Table 3 summarises the more detailed parameter set for the Shack-Hartmann and pyramid wavefront sensor models used in our yao simulations.

We analysed the performance of two slightly different SHS systems (see also Section 6.1), one with 74 sub-apertures across the pupil (SHS74) and one with 60 sub-apertures (SHS60). The detector pixel field of view (pixel scale) lies in between 1–2 pixel per size of the diffraction limited point spread function (PSF) of a sub-aperture. In our simulations we tried to set this pixel scale as close as possible to 1.2 pixel per full width at half maximum (FWHM) of the PSF. Due to the finite size and resolution of the simulation grid, we used the numbers given in Table 3. The size of the field stop matches the size of 4 pixels.

Most of the yao parameters for the modulated pyramid wavefront sensor PYR74 are the same as for the SHS74 system.

### 5.2 The yao simulation grid

The yao two dimensional simulation grid (Fig. 7) defines on how many points wavefronts are sampled and propagated. This is usually a circular area (pupil) and should have a spatial resolution that samples the spatial coherence length $r_0$ with at least 2–3 points. In the case of METIS with a near-infrared wavefront sensor

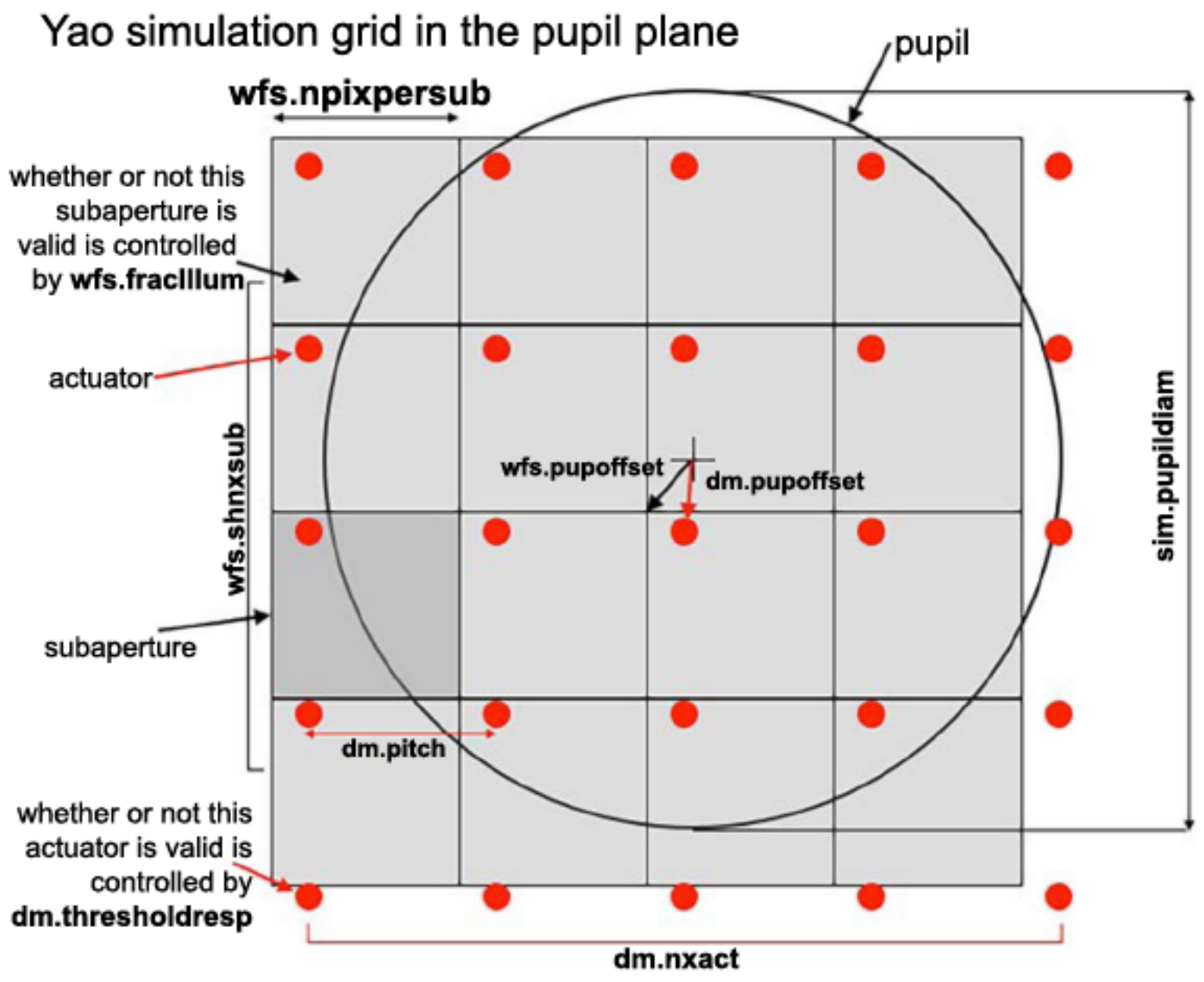

**Fig. 7** Standard geometry for a Shack-Hartmann wavefront sensor and piezo stack deformable mirror configuration used in yao simulations [30]. Parameter *sim.pupildiam* defines the diameter of the main simulation grid (pupil). Yao wavefront sensor *wfs.** and deformable mirror *dm.** parameters are explained in the text

and all science channels at wavelengths longer than 3 microns, $r_0$ is always larger than 0.3 m. We therefore selected for most simulations a grid size of $370 \times 370$ pixels, i.e. $sim.pupildiam = 370$. Mapping this grid on a 37 m telescope pupil (see Section 5.4) results in a resolution of 0.1 m per pixel, at least 3 pixel per $r_0$.

For a SHS and PYR with *wfs.shnxsub* equidistant sub-apertures over the pupil diameter, each sub-aperture is sampled with *sim.pupildiam/wfs.shnxsub* pixels on the simulation grid. Yao parameters *wfs.fracIllum* and *dm.thresholdresp* define whether a sub-aperture or actuator is used or not. In our settings we used a fractional flux limit of 0.9 (90% illuminated) for the SHS and 0.5 for the PYR simulations to define the grid of active sub-apertures. The grid of actuators can be defined using a list of x-y coordinates on the simulation grid. That is our approach when using a model of the ELT M4 deformable mirror. Alternatively, yao has an internal model for piezo stack-array deformable mirrors, defined by the number of actuators over the pupil diameter `dm.nxact` with a spacing of `dm.pitch` pixels on the simulation grid. In both cases, the not necessarily equidistant actuator grid can further be changed during calibration of the AO system using the parameter *dm.thresholdresp*. If the response of pushing an actuator during yao calibration is below this threshold with respect to the maximum response of all actuators, then this actuator is discarded. For all our simulations we set *dm.thresholdresp=0.3*. The sub-aperture grid as well as the actuator grid was centered on the pupil, i.e. `wfs.pupoffset` and

`dm.pupoffset` were set to 0. The actually used number of sub-apertures and actuators for the analysed METIS SCAO configurations are listed in Tables 3 and 5 in Section 5.7.

In our SCAO simulations we do not rotate the pupil. Although METIS has a derotator unit in the CFO (see Fig. 2), the SCAO control system will rotate the pupil numerically for certain observation modes.

### 5.3 The atmosphere and integrated turbulence parameters

For our simulations we used phase screens based on the Kolmogorov and Von Kármán model of the atmosphere [40], with an outer scale value of $L_0$ = 25 m. Our phase screens are typically rectangular with a size of 819.2 m × 409.6 m and a resolution of 0.1 m.

The optical turbulent atmosphere is described with a 35 layer model of the troposphere provided by the European Southern Observatory (ESO). For each atmospheric layer ranging from 30 m to 26.5 km above the observatory platform we can select wind speed and turbulence strength $C_n^2$ fraction for 5 different seeing conditions: median seeing, first quartile Q1 of the seeing distribution, 2nd (Q2), 3rd (Q3), and last quartile Q4 of the seeing distribution. The integrated turbulence parameters for these 5 conditions are listed in Table 4. The $C_n^2$ numbers used are based on measurements recorded on Paranal observatory which is located at a distance of about 25 km from the ELT site, using MASS-DIMM data [41]. The mean wind speeds for all seeing conditions are about 9 m/s, the highest speed of 43.84 m/s appears in Q4 for layer #19 at a height of 10500 m. For most of our simulations we used the same wind direction – parallel to the tip axes of the tip-tilt mirror – for all 35 layers. This setting has an impact on the image motion (tip/tilt) compensation because tip and tilt are directional aberrations. Tip direction is parallel to the horizontal/x-axis and tilt is parallel to the vertical/y-axis with respect to the ELT pupil mask shown in Fig. 8.

To generate the atmospheric phase screens for a simulation of a certain duration, yao only needs two parameters: the size, and the length of the outer scale.

As an example, for a 10 s simulation a minimum phase screen size of 10 * 43.84 m = 438.4 m is needed to allow yao to shift the fastest moving layer over the telescope without wrapping the phase screen.

**Table 4** Integrated turbulence parameters used in simulations

| Parameter | Q1 | Q2 | Q3 | Q4 | Median |
|---|---|---|---|---|---|
| $r_0$ [m] | 0.234 | 0.178 | 0.139 | 0.097 | 0.157 (1.35) |
| $\tau_0$ [ms] | 8.08 | 6.12 | 4.78 | 3.11 | 5.35 (45.93) |
| $\theta_0$ [arcsec] | 2.8 | 2.35 | 2.17 | 2.15 | 2.3 (19.75) |

Fried parameter $r_0$, wavefront coherence time $\tau_0$, and isoplanatic angle $\theta_0$ for a wavelength of 500 nm. Q1 to Q4 represent the four seeing quarters (quartiles), Median represents the median seeing condition. Values in parenthesis are calculated for a wavelength of 3 $\mu$m, according to the well known $\lambda^{6/5}$ scaling

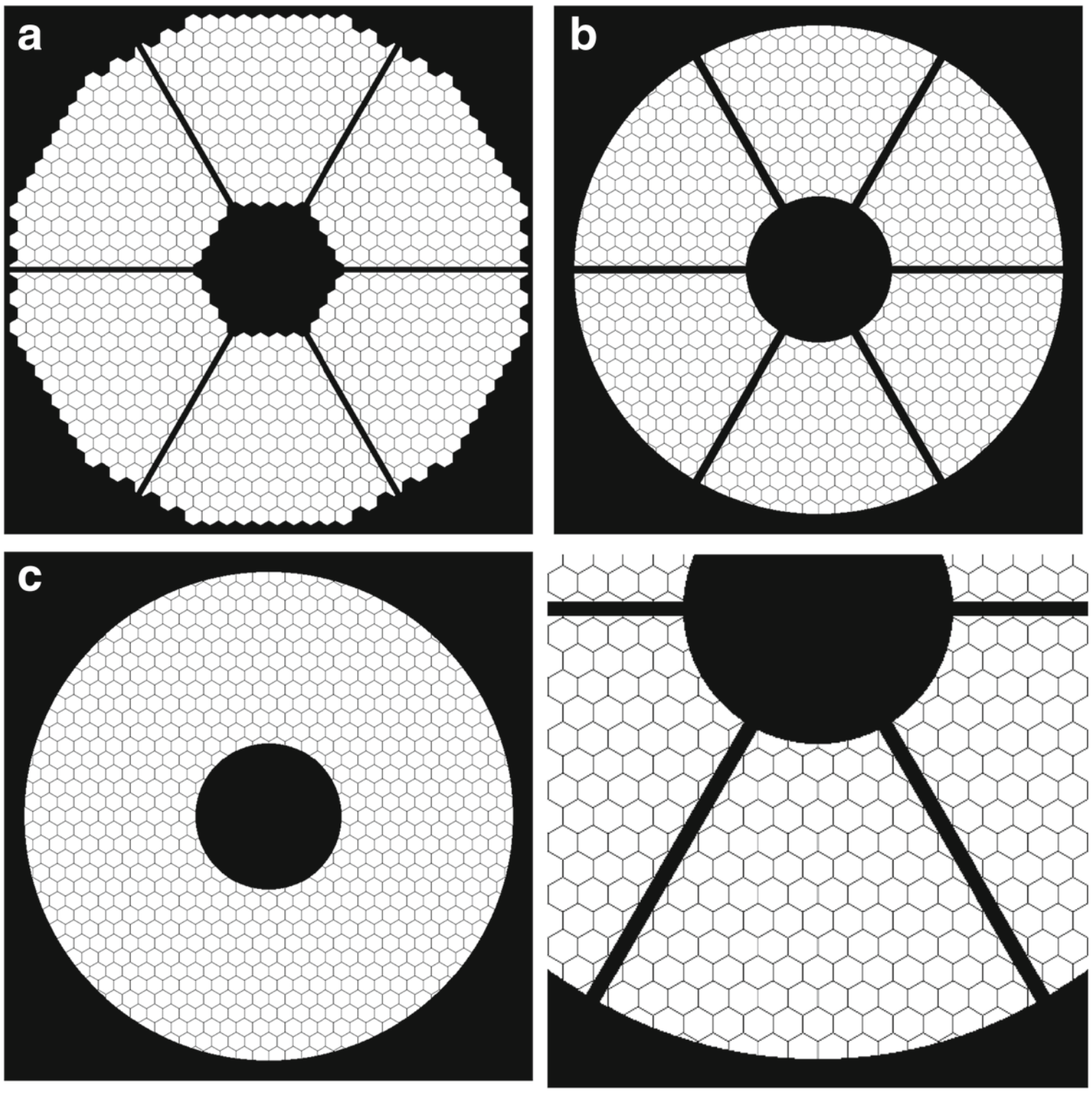

**Fig. 8** **a** 39.3 m ELT pupil with segments and overlaid with 40 cm wide spiders of the secondary mirror. **b** 37 m, circular ring masked ELT pupil with 60 cm wide spiders that we used in our PYR simulations. **c** 37 m, circular ring masked ELT pupil without spiders that we used in our SHS simulations. The lower right inset shows a zoom into the pupil area of **b** to visualise the segmented, hexagonal structure of the ELT primary mirror (M1)

### 5.4 Pupil masks and pupil fragmentation issues

In the METIS instrument design it is currently considered to use a circular, ring shaped pupil mask that clips all non-circular edges (Fig.8a) of the real ELT primary mirror (M1). This leads to a circular ring mask with an outer diameter of 37 m and an inner diameter of 11.1 m as shown in Fig.8b, c. However, the introduction of a pupil mask before the wavefront sensor is under debate. The ELT M1 has 798 hexagonal segments, each hexagonal shaped segment with a longest diagonal (tip to tip) of 1.42 m and a gap of 4 mm between neighbouring segments.

In consideration of the fact that PYR and SHS systems react differently to wavefront piston, we had to select different pupil masks for them in our simulations. The reason behind this is a recently observed phenomenon in wavefront reconstruction for AO systems with sub-aperture sizes smaller or of similar size than the width of the telescope spiders [42]. In this case, some sub-apertures are masked out such that the pupil looks fragmented from the wavefront sensor point of view. Looking at the measured wavefront slopes, gaps at or around the spider location can lead to the reconstruction of a differential piston between the fragments. With the available reconstruction methods in yao, only the pyramid wavefront sensor was able to reconstruct a wavefront over the whole pupil that did not show piston offsets between fragments. The effect is demonstrated in Fig. 9. Here, we deliberately feed the system with a wavefront that has a large piston offset (800 nm) applied to one of the fragments. Running each system in closed-loop for 80 ms (80 correction cycles), we find that the pyramid WFS is able to correctly reconstruct the piston offset and correct the aberration down to a flat wavefront, whilst the Shack-Hartmann sensor is partially blind to the piston and cannot fully correct it, leaving a residual offset of about 430 nm between the fragment concerned and the remainder of the pupil.

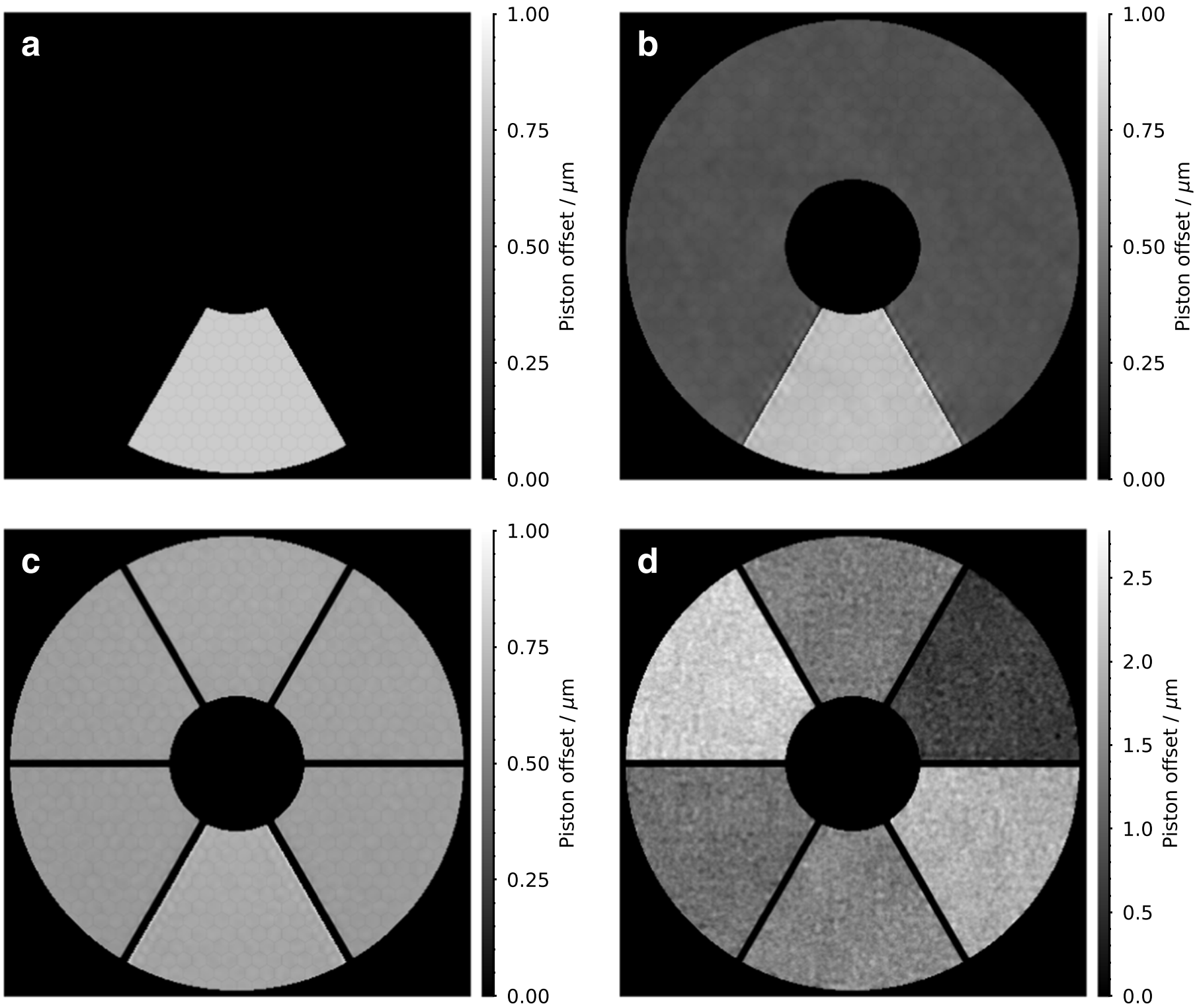

**Fig. 9** **a** – **c**): comparison between a Shack-Hartmann (SHS74) and Pyramid (PYR74) wavefront sensor based AO system with respect to a pure 800 nm piston applied to the six o'clock fragment of the ELT pupil (**a**). **b**: residual phase produced with SHS74 after running in closed-loop for 80 ms. **c**: residual phase produced with a PYR74 after running in closed-loop operation for 80 ms. For **a**, **b** and **c** no atmosphere turbulence was considered. **d** same as **b** obtained under median seeing conditions without additional piston applied to the six o'clock fragment. To distinguish the piston effect from the pupil fragmentation effect for SHS74, we used the ELT pupil without overlaid spiders for **b**. Details of the PYR74 and SHS74 configurations are given in Section 5.1

In turn what we are seeing is that in regular closed-loop operation, i.e. without artificial piston terms applied, the SHS frequently erroneously reconstructs a piston term for one or more fragments. Also the pyramid WFS can show this behavior, but in our simulations significantly less often. These terms then pile up over the course of the loop cycles and produce patterns similar to the one shown in Fig. 9d. Such a pattern in turn leads to PSF aberrations like the ones first observed in SPHERE [43] which were initially dubbed "Mickey Mouse effect" because of the shape of the aberrated PSF, and which is now known under the term "low-wind effect" (sometimes also named island effect).

In fact there are two effects, an intrinsic one, due to spiders, the resulting fragmented pupil, and erroneous wavefront reconstruction. And, there is an extrinsic effect that can create differential wavefront piston among the fragments due to thermal effects, i.e. temperature differences between the spider structure and the ambient air. The latter effect is the eponym for the term low-wind effect, as under low-wind conditions there is no thermal equilibrium around the spider structure during nightly observations.

It has been stated several times in the literature [44] that classical SHS systems cannot sense piston, at least when operated in the usual way with slopes derived from centroids. In contrast, pyramid systems are renown for being able to sense both slopes and phases, depending on modulation amplitude [45].

We do not want to go into details about this issue of differential piston effects for an AO system at the ELT. This will be addressed in a future publication. For the time being, we use different pupil masks for SHS and PYR systems in order to bring the SHS performance to the same level with respect to differential piston as the PYR system.

### 5.5 Non-common path aberrations

Non-common path aberrations appear after the point where science channel and wavefront sensing channel separate. For METIS this point is implemented using a dichroic mirror that reflects the star light with a wavelength $< 3\mu m$ to the wavefront sensor and transmits all light with longer wavelengths to the science channel. Optical aberrations that appear only in the science channel can be measured and transferred to the SCAO system for proper processing and correction. Section 6.3 gives more details and discusses the corresponding simulation results.

### 5.6 Wind induced vibrations

To simulate dynamical errors induced by wind on the ELT's main structure and secondary mirror unit, ESO provided 5 min long time series of tip and tilt (in arcsec on sky), which we can load into yao. The power spectrum of these perturbations (Fig. 10) shows that most energy is between 0.1 – 1 Hz and drops to zero beyond 20 Hz. In our simulations, tip and tilt time series are added step by step to the actual tip-tilt mirror shape in closed-loop iteration. In this study, we only considered vibrations induced on the ELT secondary mirror.

### 5.7 Deformable mirror models used

Yao comes with integrated deformable mirror (DM) models, such as piezo stack array DM, bimorph DM, various DMs described with 3-dimensional functions such as

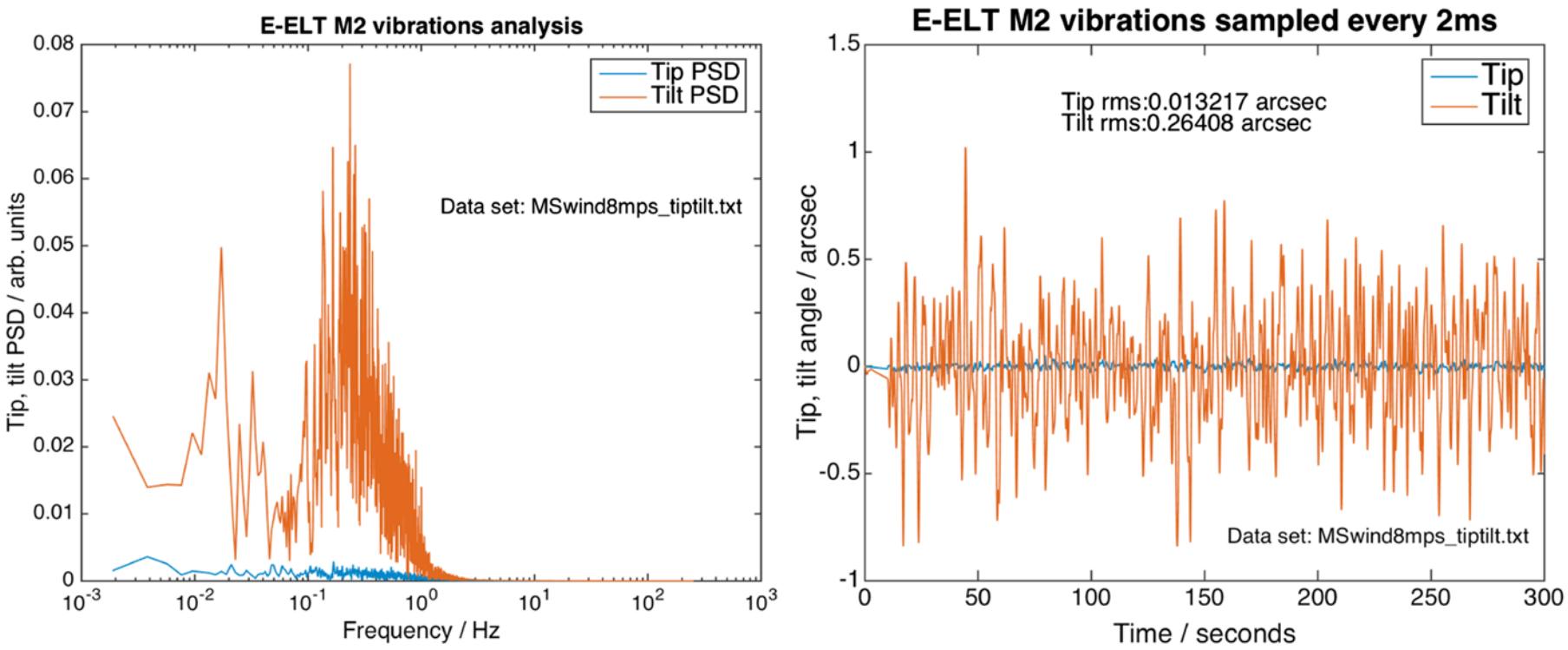

**Fig. 10** Left: Power spectral density of the wind induced vibrations on the ELT's secondary mirror. Right: 5 min sequence of the vibrations amplitude on the ELT's secondary mirror. In both plots 2 directions, tip (blue) and tilt (red) are shown

Zernike polynomials, disk harmonic, or Karhunen-Loève functions, segmented DM, and user defined DM. Here we use either the yao internal stack array model, or our own ELT M4 model. In both cases influence functions at certain positions (actuator locations) are used to compute the DM shape. Either the influence functions are given over the whole simulation grid or only a local version around the actuator location is used. For the ELT DM the latter case consists of a set of 5316 influence functions, with each influence function described on a $40 \times 40$ pixel grid (Fig. 11 right). For the stack array DM this grid has a size of $24 \times 24$ pixels (Fig. 11 left). Using small sized influence functions saves computer memory and reduces computation time. The actuator spacing for both models corresponds to 0.5 m on the ELT primary mirror. The actuator grid is equidistant for the yao stack array model and non-equidistant for the ELT M4 model.

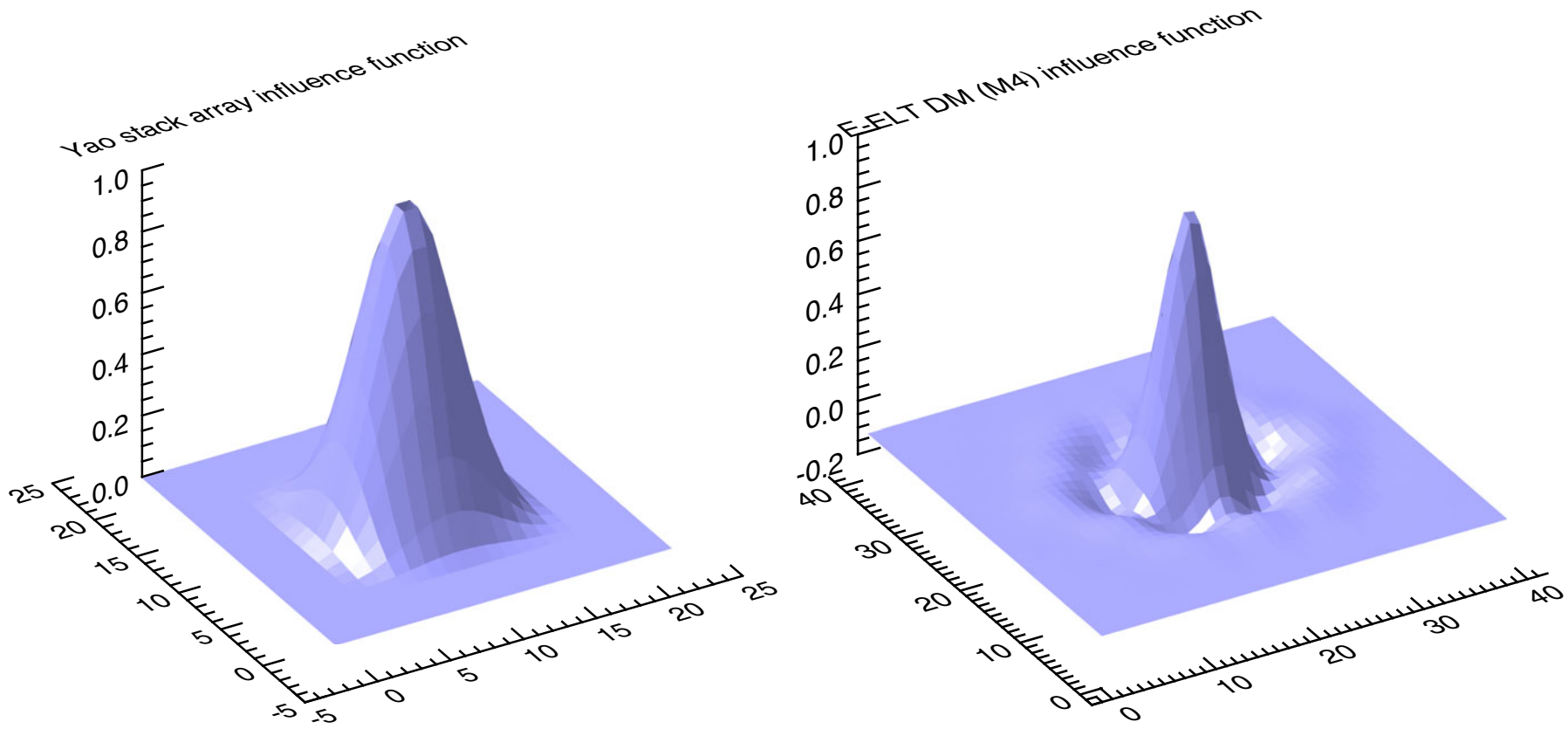

**Fig. 11** Left: standard influence function in units of microns of one actuator of a yao stack array deformable mirror on a $24 \times 24$ pixel grid. Right: typical influence function of one actuator of the ELT deformable mirror model on a $40 \times 40$ pixel grid. Note the negative deformation of the ELT actuator. Regardless of this, if all actuators are pushed with the same value/voltage, the result over the yao simulation grid gives a sufficiently flat surface. Actuators outside the simulation grid are also used to avoid edge effects

When we use the stack array DM in our simulations, the number of equidistant actuators over the pupil diameter matches the number of linear sub-apertures plus one. That means the geometry or the registration of sub-apertures and actuators follows the Fried geometry (actuators are located at the corners of a sub-aperture) as shown in Figs. 7 and 12. For the ELT M4 model, the geometry deviates slightly from the Fried geometry as shown in Fig. 12 on the right.

The actual number of actuators used in closed-loop operation differs slightly for each wavefront sensor in use as explained in Section 5.2. The number of active actuators after calibration for each wavefront sensor is listed in Table 5. For all METIS SCAO simulations, a tip-tilt mirror with 2 actuators has been used. During calibration this mirror was actuated either with an angle equivalent to 200 mas, 50 mas or 20 mas on sky.

Note that the METIS SCAO wavefront sensor is conjugated to the ELT M4 deformable mirror. The tilted ELT M4 itself is conjugated to the atmospheric ground-layer around 530–630 m above the telescope. In our simulations however, the high-order DM as well as the low-order tip-tilt mirror are conjugated to the ground (0 m). Additionally, the ELT primary mirror M1, which defines our pupil and the corresponding pupil masks used in our simulations are in reality out of focus from the WFS point of view. Both effects will be addressed in our end-to-end simulations planned for 2019/2020.

## 5.8 Control loop parameters

First of all it should be mentioned that the ELT control system, especially the control of the ELT M4 DM, does not allow control frequencies higher than 1 kHz. To measure

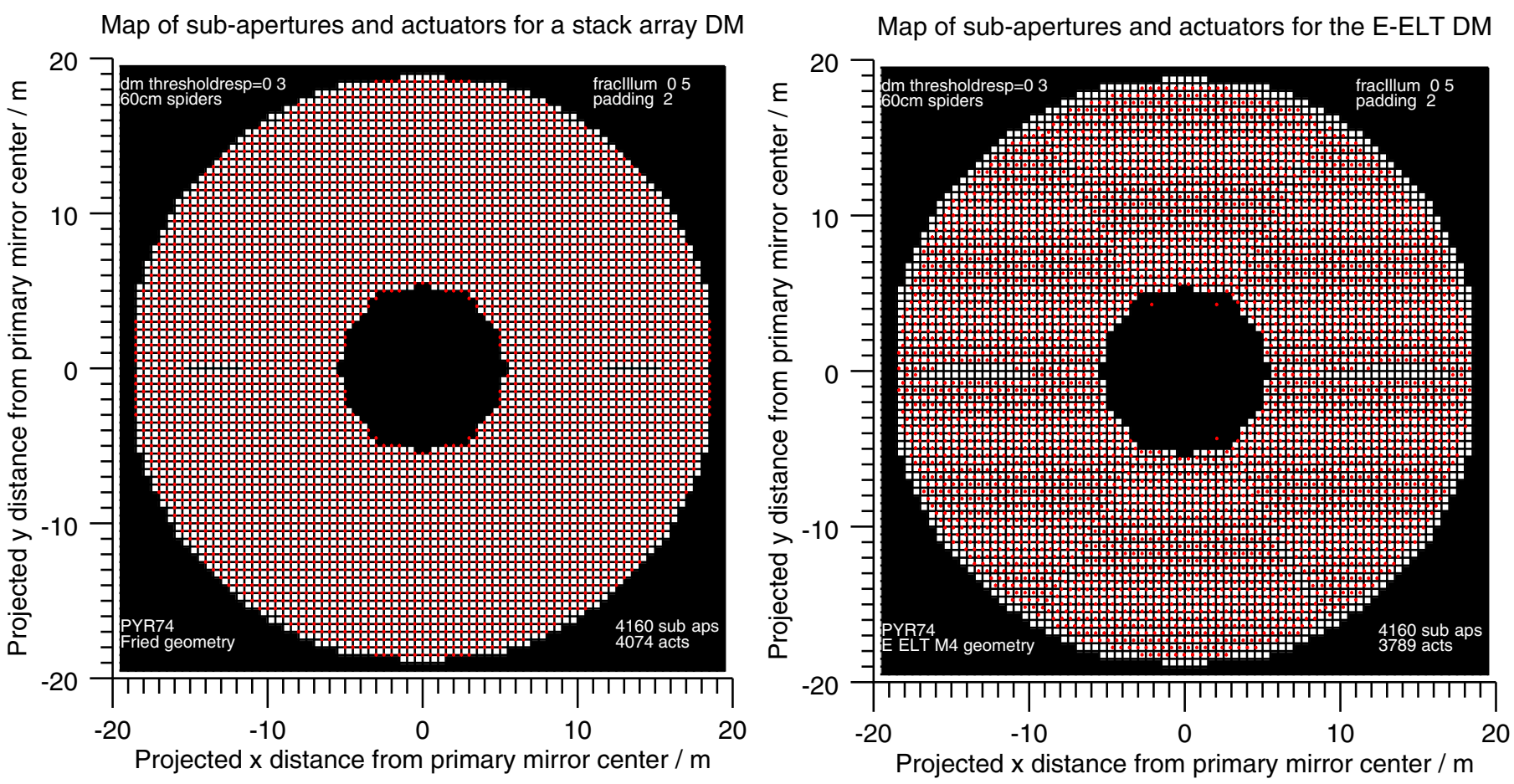

**Fig. 12** Left: sub-aperture and actuator (red dots) registration map for PYR74 and stack array DM model. Right: sub-aperture and actuator registration map for PYR74 and the ELT M4 model

**Table 5** Number of active actuators for 2 deformable mirror models and various wavefront sensor configurations (SHS60, SHS74, PYR74)

| Deformable mirror model | SHS60 | SHS74 | PYR74 |
|---|---|---|---|
| Stack array | 4036 (4601) | 4008 (4601) | 4074 (4601) |
| ELT M4 | — | — | 3789 (4012) |

Numbers in parenthesis give the number of actuators used during calibration. After calibration some actuators are discarded depending on the sensitivity of the wavefront sensor to them. This behaviour is controlled with the yao parameter `dm.thresholdresp=0.3`. See Section 5.2. The ELT M4 model was only used for PYR simulations the wavefronts sufficiently fast, the rule of thumb is to sample the wavefront about 15–20 times faster than the coherence time. Our choice of sampling frequencies fits well with a median coherence time at 3 $\mu$m of 46 ms (see Table 4).

Therefore, for all our simulations we use a loop delay of either 2 ms for the 1 kHz AO loop frequency or 4 ms for the 500 Hz loop frequency. That means the actual wavefront measurement at iteration $n$ is used at iteration $n + 2$. This behaviour is controlled with the yao parameter *loop.framedelay* = *2*. After reconstruction of the wavefront, the corresponding DM shape $DM_{err}$ is calculated. Using a proportional integral (PI) control law, $DM_{err}$ is multiplied with the gain factor for this DM and the result subtracted from the DM shape of the previous iteration. The product of the yao parameters *loop.gain* and *dm.gain* sets the individual gain factor for each DM. One goal of our simulations is to optimise this gain factor. This gain factor is, in the general case, only the *loop.gain*.

## 6 Wavefront sensor selection

### 6.1 Test cases for the selection process

The test cases we used for our simulations are summarised in Table 6. The type of wavefront sensor and the wavefront spatial sampling for each wavefront sensor are summarized in Table 2.

We simulated 5 seeing conditions (see Section 5.3), 0.43″, 0.57″, 0.64″, 0.73″, 1.04″, at 3 zenith distances of 0, 30, and 60 degree. The brightness of the natural guide star was varied between $m_K = 2.8$ mag (very bright), $m_K = 7.1$ mag (bright), and $m_K = 10.35$ mag (faint). Additionally, a guide star with $m_K = 10$ mag was simulated to check the requirements listed in Section 2. The brightness magnitudes where chosen to match a detected flux of approximately 5000, 100, and 5 electrons per sub-aperture and millisecond (see Table 6). Two AO loop frequencies, 500 Hz and 1000 Hz complete the parameter space we want to explore.

Although the ELT control system limits the control frequency for the M4 DM to 1 kHz, we have also carried out a simulation with an AO loop frequency of 2 kHz. This leads to a significant improvement in performance in contrast in the context of high-contrast imaging around very bright stars [6].

**Table 6** Simulation parameter space summary

| Parameter | Values | Class | Comment |
|---|---|---|---|
| Wavefront sensor | SHS60, SHS74, PYR74 | explore | see Section 5.1 |
| Guide star magnitude | 2.8, 7.1, 10.0, 10.35 | explore | Stellar magnitude in K-band (2.2$\mu$m) |
| Detected signal | 5000, 100, 6.7, 5 | | Approximate WFS signal in electrons/sub-aperture/millisecond |
| Background flux | 12.9 | fixed | magnitude/arcsec$^2$ in K-band (2.2$\mu$m), including sky and telescope emissivity |
| Seeing | 0.43, 0.57, 0.64, 0.73, 1.04 | explore | Seeing at $\lambda = 500$ nm in arcsec |
| AO loop frequency | 500, 1000 | explore | Hz |
| Zenith distance | 0, 30, 60 | explore | degree |
| Control loop gain | 0.17, 0.35 | optimise | see Section 5.8 |
| Regularisation parameter | 0.01, 0.025, 0.05, 0.1 | optimise | control matrix creation, see Section 4.1, (3) |
| Pixel threshold | 0, 1, 3 | optimise | electrons/pixel, see Table 3 |
| Iterations | 2200 | fixed | total number of iterations per simulation |
| Vibrations | 0, 25%, 100% | fixed | according to Section 5.6 |
| Non-common path aberrations (NCPA) | 0-400 nm rms | explore | see Section 6.3 |

The detected signal values consider an atmospheric transmission of 0.9, a telescope/instrument transmission of 0.25, and a detector quantum efficiency of 0.7

The product of parameters to explore and optimise calls for total 8640 yao simulations (2880 per sensor type), each running 2200 iterations. On our Dell Poweredge R830 hardware with 88 CPUs running at 2.2 GHz, one iteration takes about 1 second/CPU for the SHS60/SHS74 simulations and about 1.5 s/CPU for the PYR74 simulations. With 2 R830 systems available, the full suite of 8640 simulations took less than 2 days.

### 6.2 Results of the selection process test cases

The Strehl ratio plots shown in Figs. 13, 14, 15, 16, and 17 are horizontal lines for guide star magnitudes up to a guide star magnitude of $K \approx 7$. This high flux regime is dominated by the deformable mirror fitting error. Towards fainter guide stars, the AO performance drops because the low photon flux reduces the quality of the reconstructed wavefront. In particular the SHS74 system cannot fulfil the requirement of R-MET-111 (Section 2) for a loop frequency of 1000 Hz. Using a loop frequency of 500 Hz, SHS74 meets the requirement but not the goal of R-MET-111 (Fig. 14 center, Fig. 16 center). Only PYR74 surpasses the goal requirement. All test cases comply to the R-MET-119 (Section 2) requirement. The Strehl ratio plots further show the superiority of the PYR74 system in all cases, clearly visible for fainter reference sources.

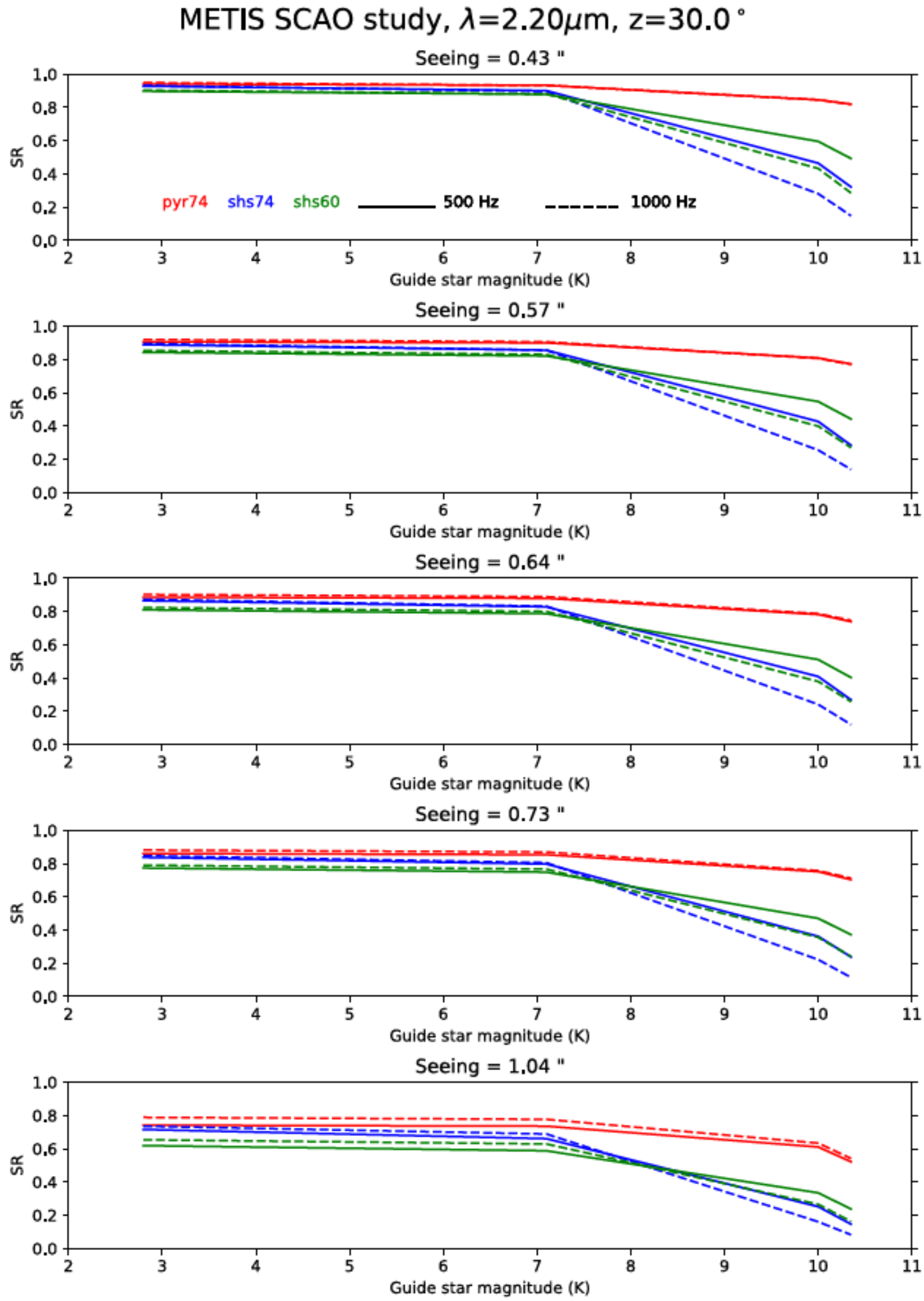

**Fig. 13** Strehl ratio (SR) at the wavefront sensor wavelength of 2.2$\mu$m as a function of guide star magnitude in K-band for five seeing conditions and a zenith distance of z = 30°. Full lines show the 500 Hz AO loop frequency results, dashed lines the 1000 Hz results. Strehl ratios have only been evaluated for guide star magnitudes of K = 2.8, 7.1, 10.0, and 10.35. Lines are only shown to guide the eye. Red colour lines show the results for the pyramid wavefront sensor with 74 equally spaced sub-apertures (PYR74) across the 37 m ELT pupil. Blue coloured lines represent the Shack-Hartmann wavefront sensor corresponding with 74 sub-apertures (SHS74), and green coloured lines represent the Shack-Hartmann wavefront sensor corresponding with 60 sub-apertures (SHS60). For each Strehl ratio, the regularisation parameter (3), the AO/DM loop gain (Section 5.8), and the detector pixel threshold (Section 5.1) are selected such that SR is highest

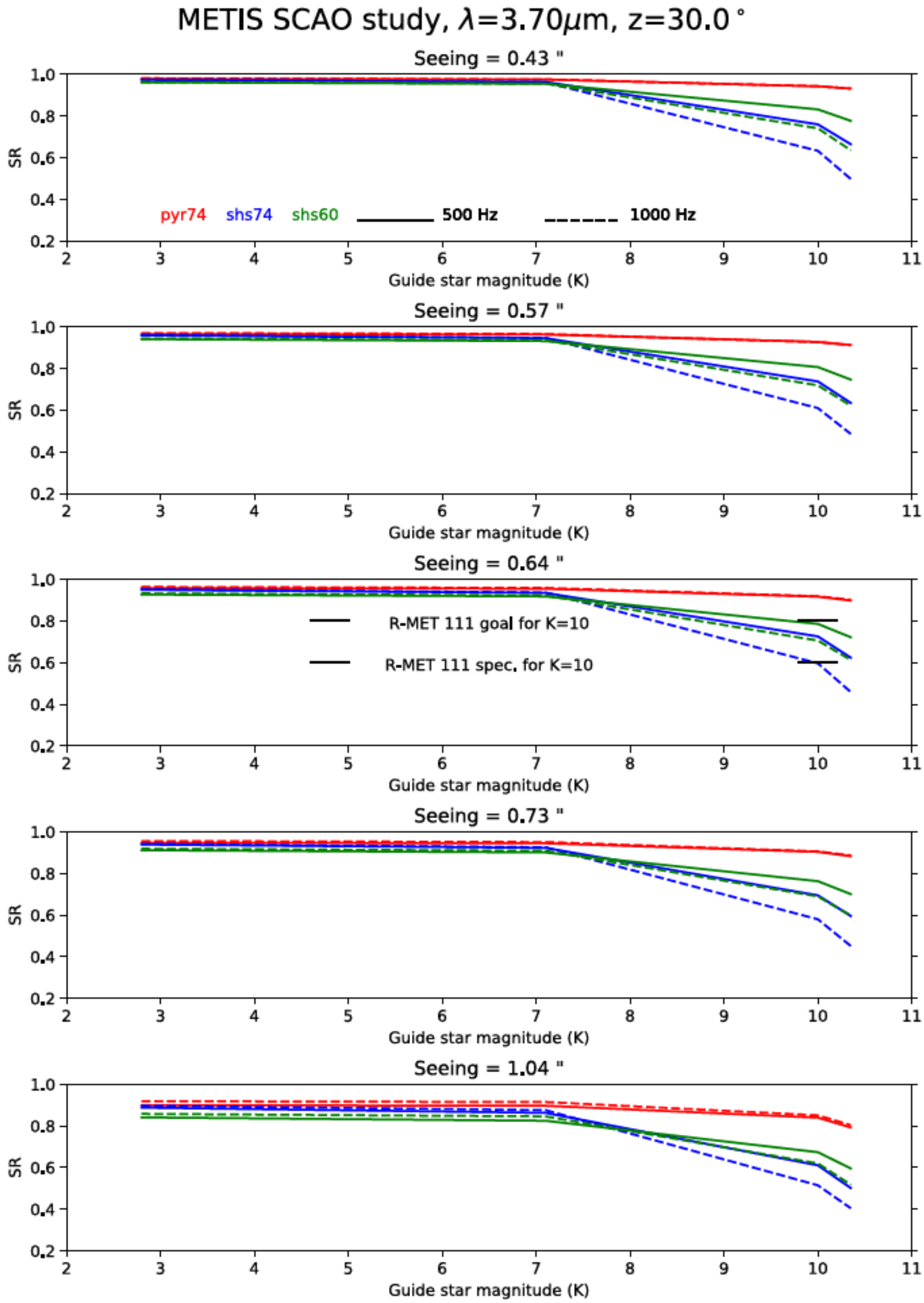

**Fig. 14** The same as Fig. 13 for the METIS science channel wavelength of 3.7$\mu$m. In addition, the center plot shows the minimum required Strehl ratios according to the requirement R-MET-119 (Section 2). The SHS74 system barely achieves the requirement but misses the goal

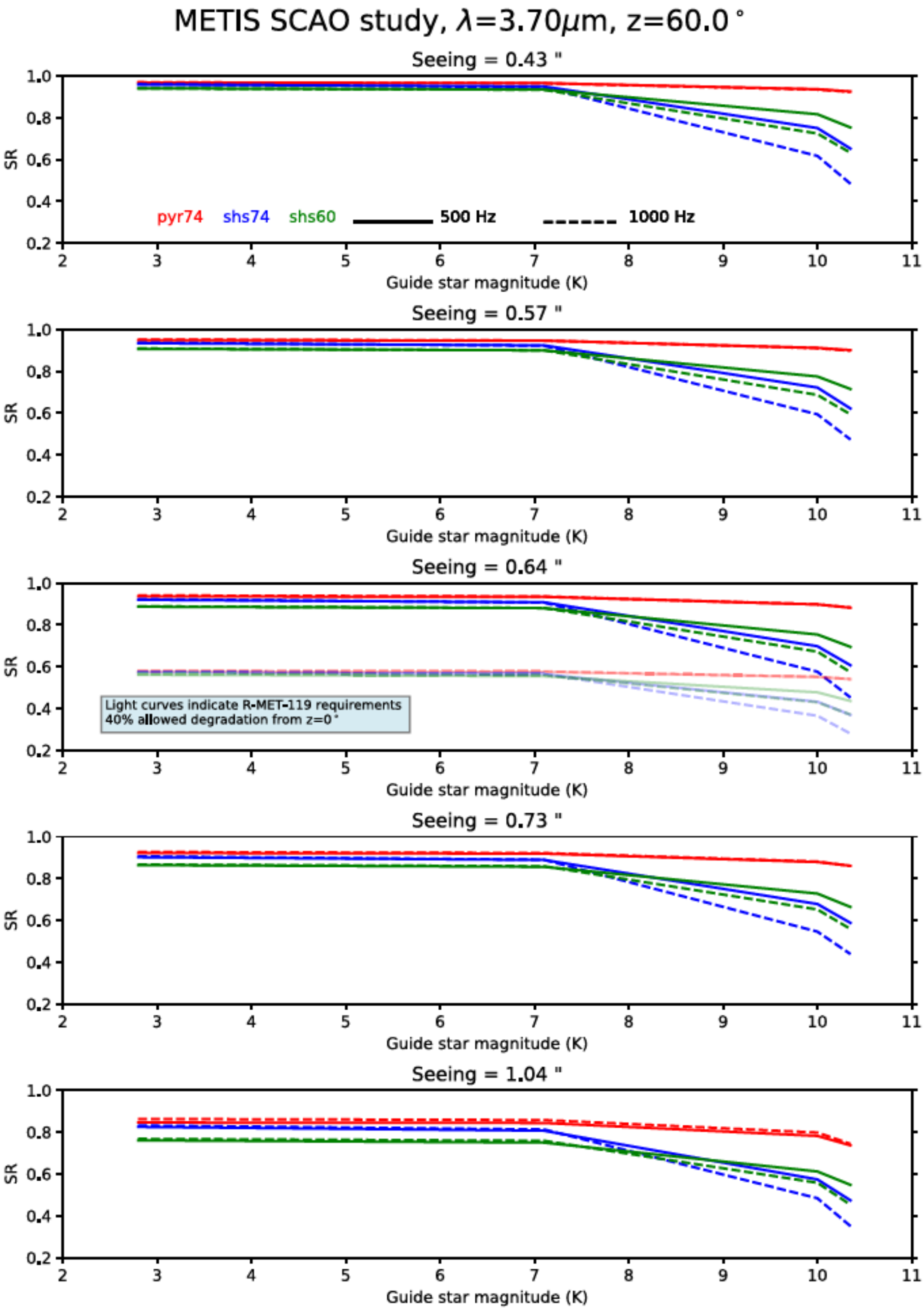

**Fig. 15** The same as Fig. 13 for the METIS science channel wavelength of 3.7$\mu$m and a zenith distance of z = 60°. In addition, the center plot shows the minimum required Strehl ratios according to the requirement R-MET-119 (Section 2)

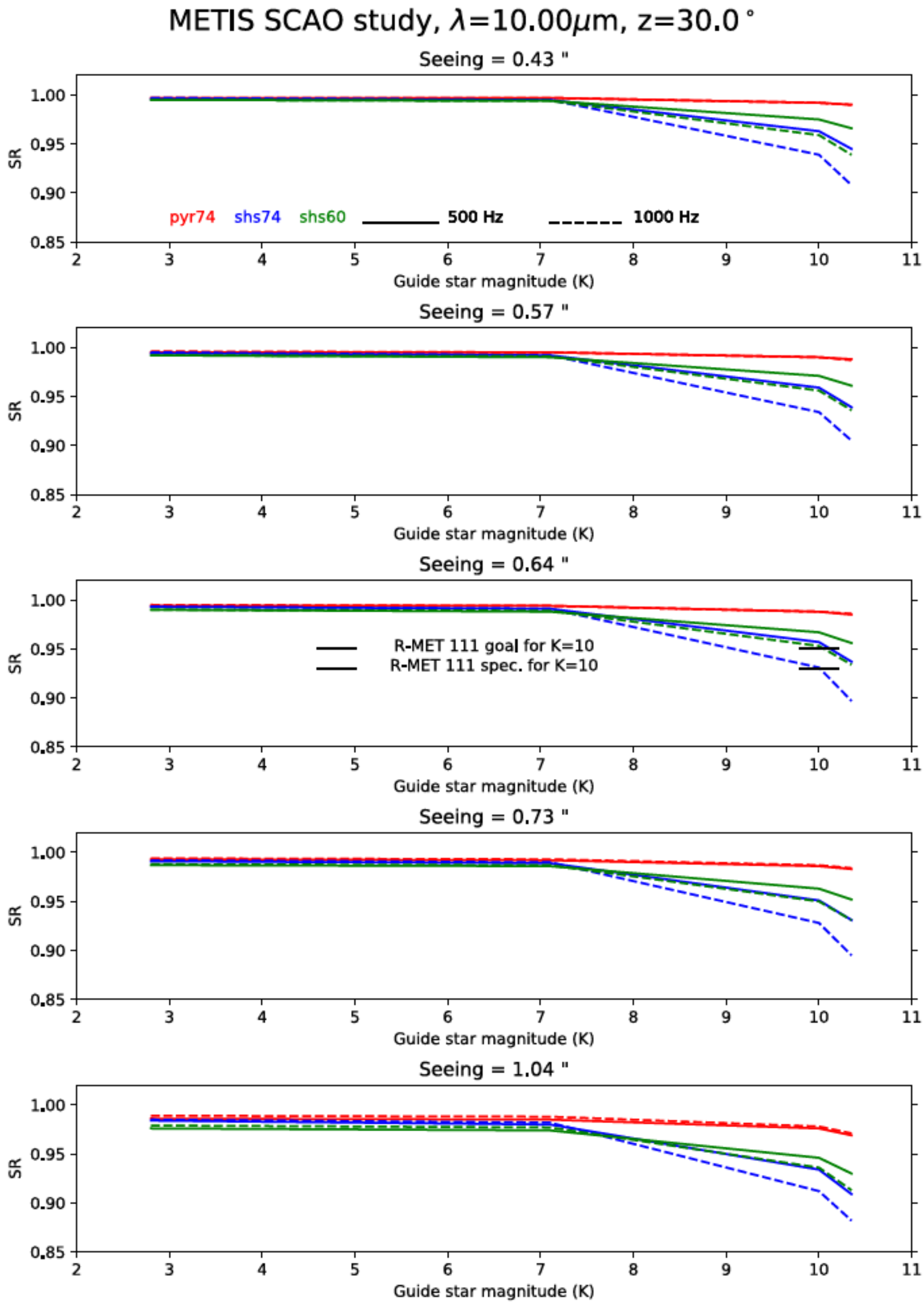

**Fig. 16** The same as Fig. 13 for the METIS science channel wavelength of 10.0$\mu$m. In addition, the center plot shows the minimum required Strehl ratios according to the requirement R-MET-119 (Section 2). The SHS74 system barely achieves the requirement but misses the goal

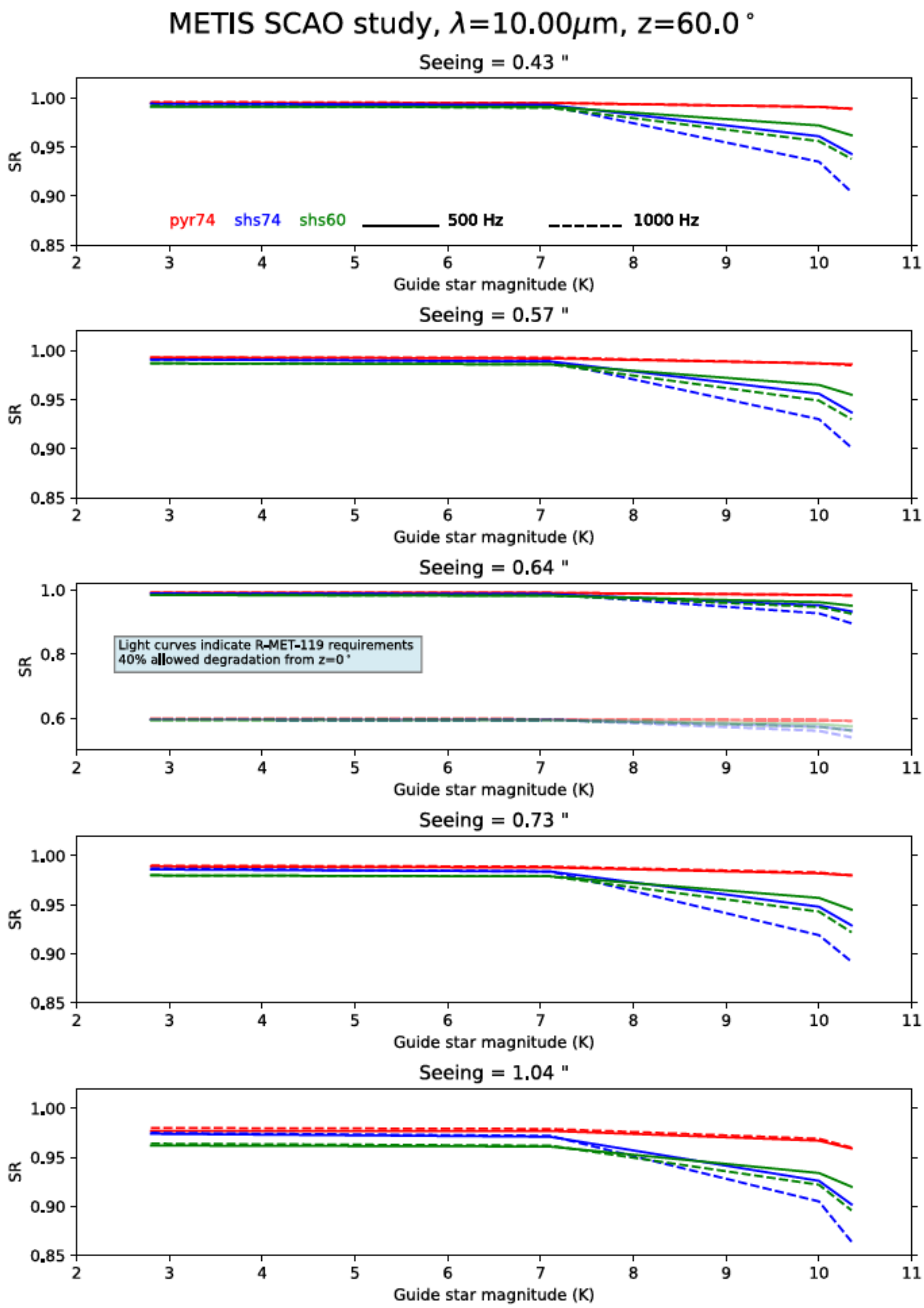

**Fig. 17** The same as Fig. 13 for the METIS science channel wavelength of 10.0$\mu$m and a zenith distance of z=60°. In addition, the center plot shows the minimum required Strehl ratios according to the requirement R-MET-119 (Section 2). Note the y-axis limits

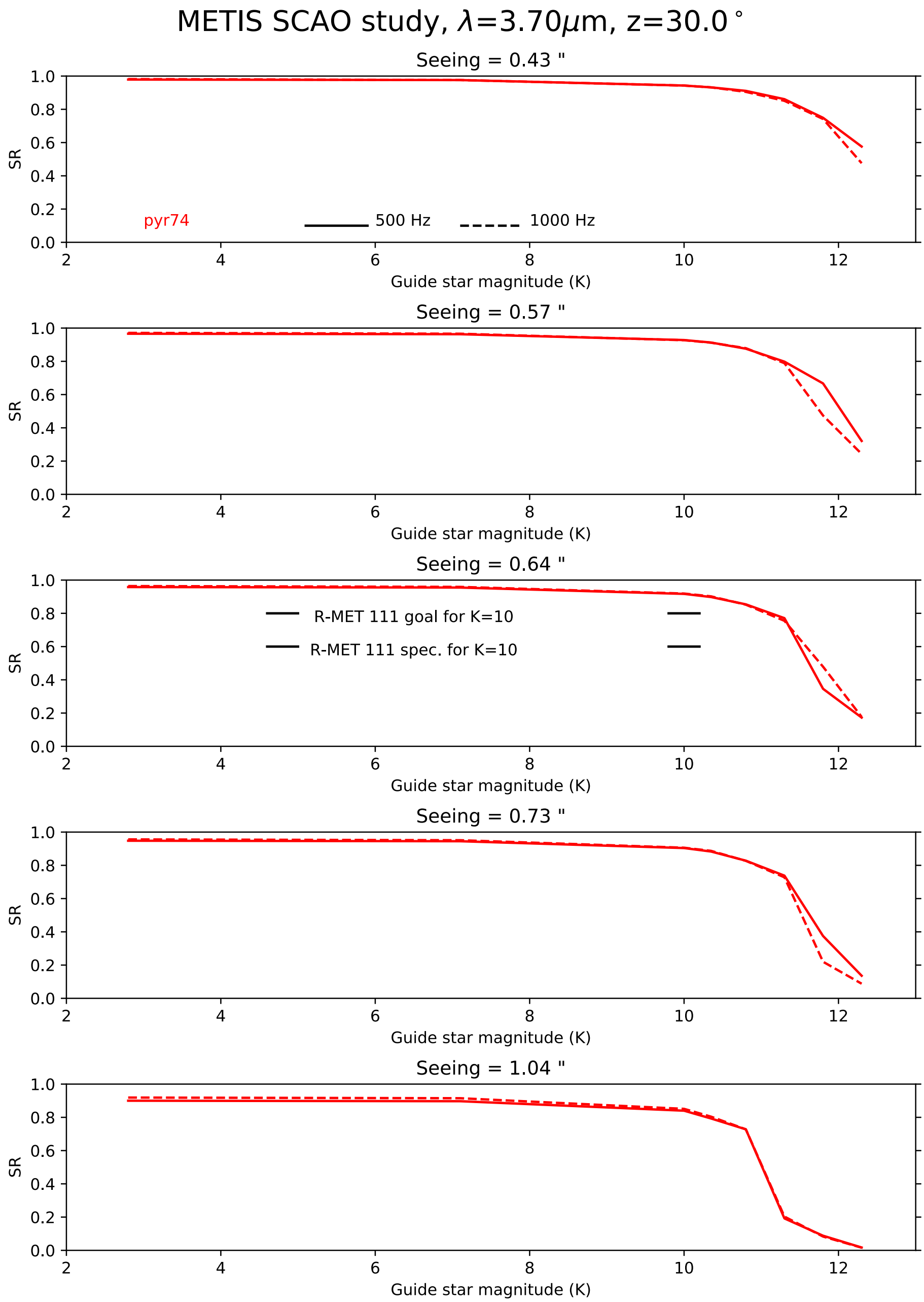

**Fig. 18** The same as Fig. 13 for the METIS science channel wavelength of 3.7$\mu$m and fainter guide star magnitudes for the PYR74 WFS only. In addition, the center plot shows the minimum required Strehl ratios according to the requirement R-MET-119 (Section 2)

With these results we concluded the selection process choosing the PYR74 system as baseline for all further investigations and discussions below.

Similar results, showing advantages of a pyramid vs. a Shack-Hartmann system have been found by other authors as well, see for example [46]. Figure 18 confirms this behaviour again if we look at the AO performance of the PYR74 system for even fainter reference stars. Another key criterion for the selection process was of course the inability of the SHS systems to cope with the fragmented pupil at all.

The high contrast imaging channel of METIS requires that the residual image motion stays below 5 mas rms (Section 2). In our analysis we use the recorded residual wavefronts of yao and decompose them into 10 Karhunen-Loève modes including central obscuration. We use routines originally written and provided by R. Cannon [47] and convert the resulting modal coefficients obtained for tip and tilt to angles on sky.

Figures 19 and 20 show the residual image motion recorded over 2.2 s for the PYR74 system without vibrations, and recorded over 23 s with wind induced vibrations on the ELT secondary mirror (M2) as shown in Section 5.6, Fig. 10 starting at second 10. The conditions for these simulations are median seeing and a guide star magnitude of K = 7.1.

As can be seen in Fig. 20, the residual tilt amplitude does not exceed the 5 mas rms requirement. We expect that even better vibration compensation including predictive control is possible [48–50].

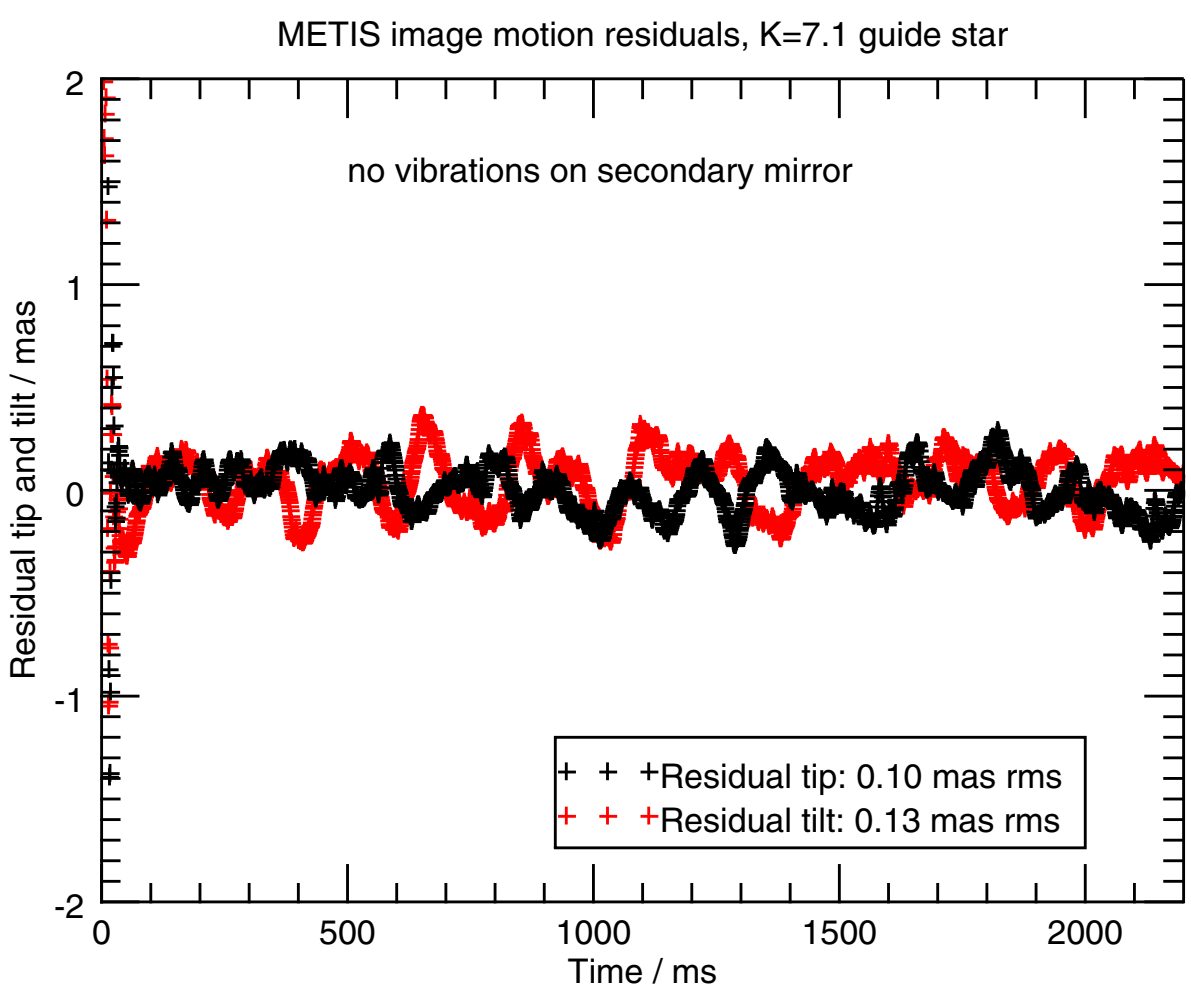

**Fig. 19** Residual image motion obtained with the PYR74 system and a guide star magnitude of K = 7.1. No vibrations have been applied to the ELT secondary mirror. The rms values are calculated 100 ms after start. Calibration parameters: high-order DM actuator calibration amplitude = 250 nm, tip-tilt mirror calibration amplitude = 50 mas. Control matrix regularization parameter=0.1. Loop parameters: high-order DM gain = 0.45, tip-tilt mirror gain = 0.54, detector pixel threshold = 1. Random wind directions

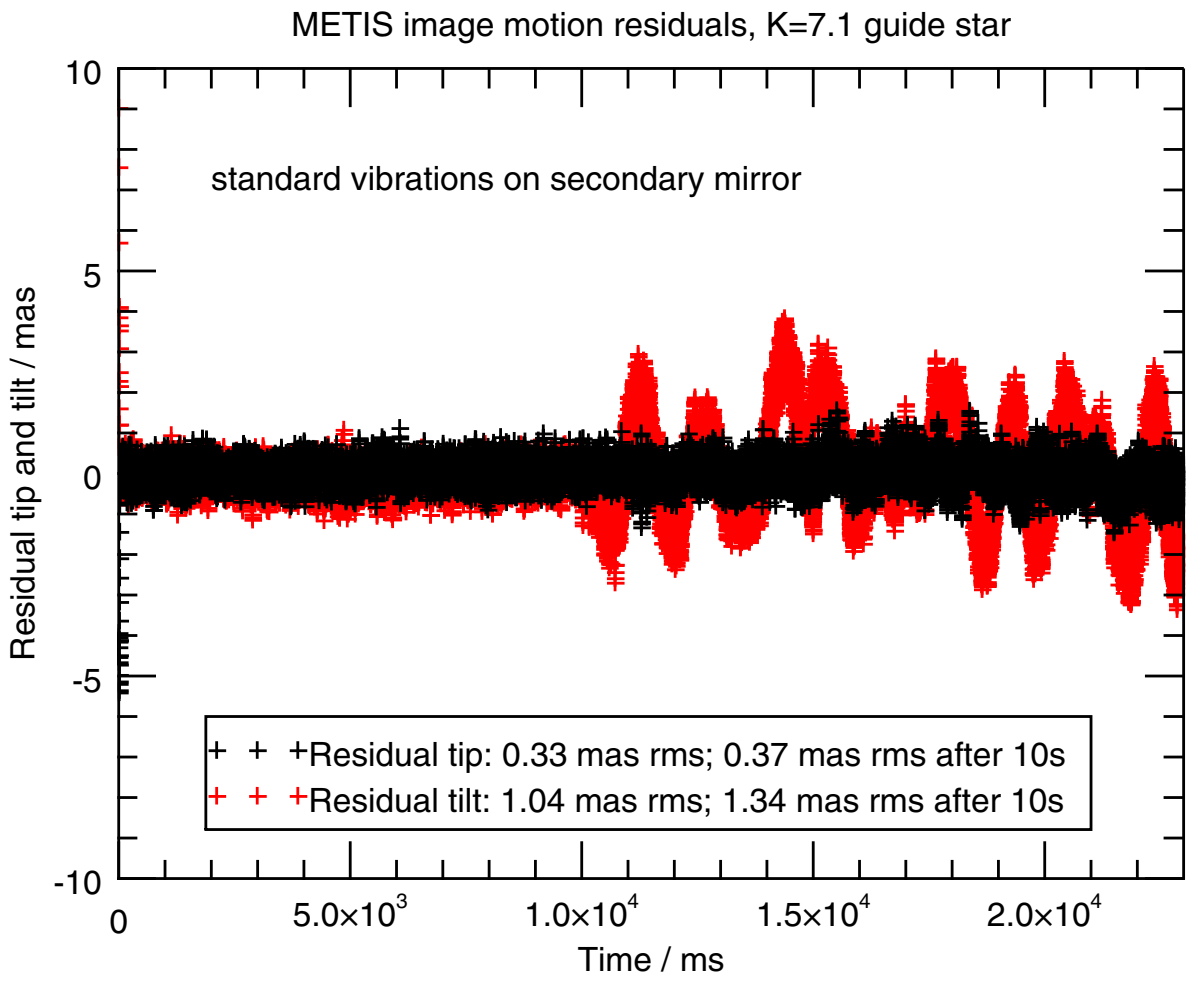

**Fig. 20** Same as Fig. 19 over 23 s with wind induced vibrations applied to the ELT secondary mirror. The vibration spectrum is shown in Fig. 10 with the first 10 s skipped. The rms values are calculated 100 ms and 10 s after start. The increase of the tilt residuals after around 10 s indicates that the AO loop control parameters need further optimization to better compensate wind induced vibrations

### 6.3 Results including non-common path aberrations

Evaluating the performance of the PYR74 system with the inclusion of the correction of non-common path aberrations is an important task as the results have an impact on the instrument design, in particular the error budget of the optical design. Non-common path aberrations (NCPAs) are typically all optical aberrations that may appear in the instrument after the position where the light of the telescope is split into the science channel and wavefront sensor channel. Looking at Fig. 2 this split position for METIS is labelled CFO-AFP, located inside the common fore optics box.

The current optical design of METIS as shown in Fig. 2 foresees non-common path aberrations between the pyramid focal plane (SCA-FP2) and the science focal plane of the LM-band imager (IMG-LM-FP1) of 210 nm rms. Since moving components such as the de-rotator and ADC are in common path, we accept the NCPAs as static aberrations.

In our current design we plan to retrieve NCPAs using the phase-sorting interferometry technique as described in [51].

In our simulations we add NCPAs to the science channel only and tell the wavefront sensor to correct them using reference slope offsets. The result of this scheme is that the SCAO system will apply a bias shape on the deformable mirror and the wavefront sensor will always “see” this bias while the science channel will benefit from the NCPA correction. In our NCPA simulations and following Noll’s ordering scheme [52], we used the Zernike modes 4 to 9 to model the NCPAs. The resulting phase maps are added to the science channel using the optics structure interface available in yao. To check whether deviations from our so far fixed pyramid beam modulation amplitude of 4λ/D can improve the SCAO performance when NCPAs

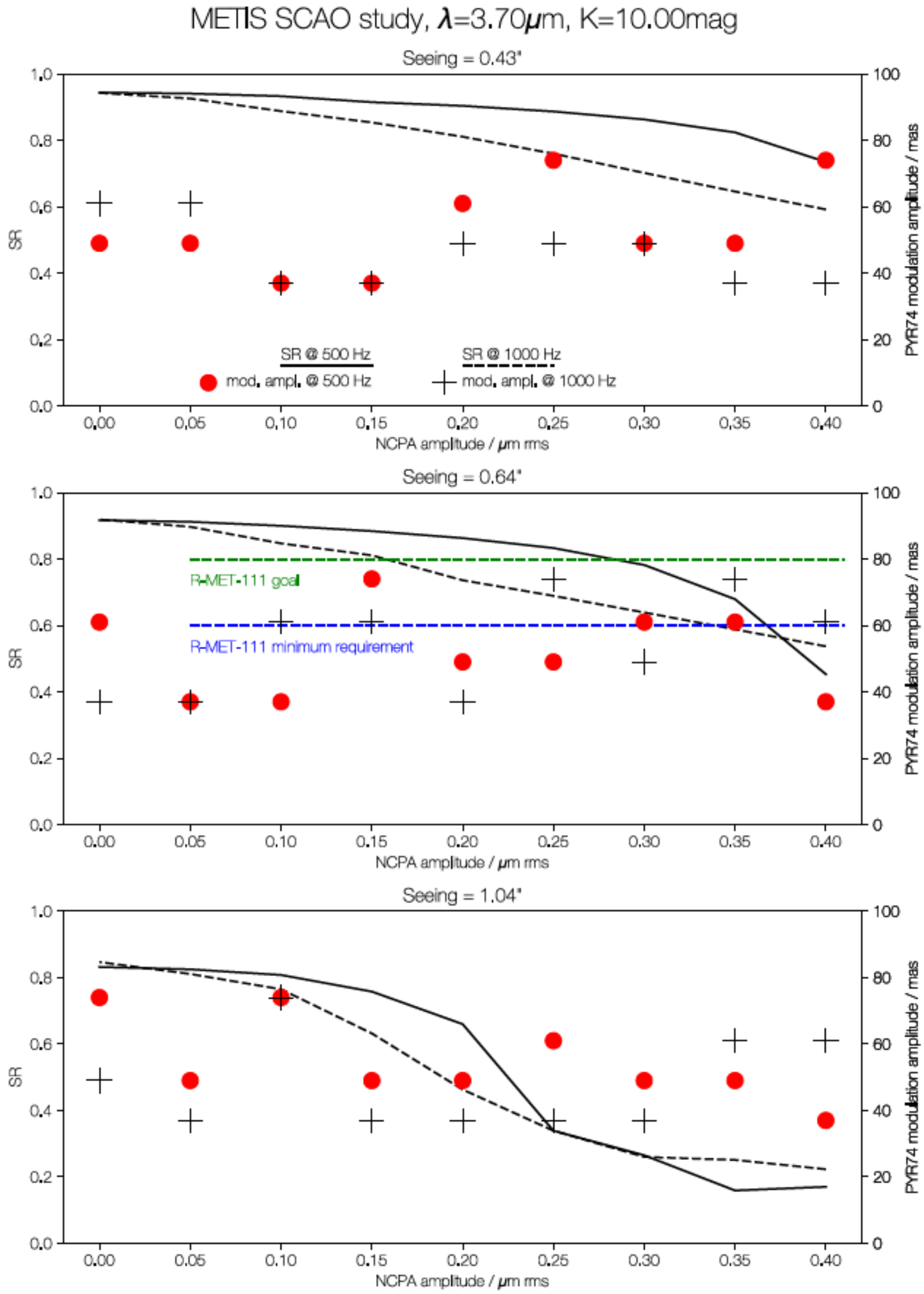

**Fig. 21** Strehl ratio (SR) at the METIS science channel wavelength of 3.7$\mu$m as a function of NCPA amplitude for 3 seeing conditions, a guide star K-band magnitude of K = 10, and a zenith distance of z = 30°. Full lines show the 500 Hz AO loop frequency results, dashed lines the 1000 Hz results. Lines are only shown to guide the eye. For each Strehl ratio, the regularisation parameter (3), the pyramid beam modulation amplitude, and the detector pixel threshold (Section 5.1) are selected such that SR is highest (see Table 6 for the used parameter space). The AO/DM loop gain was fixed to 0.17. The minimum requirement of R-MET-111(Section 2) as well as the goal requirement are shown in the center plot. The selected beam modulation amplitudes are shown as circles for the 500 Hz case and as crosses for the 1000 Hz case. Four modulation amplitudes could be selected: 3, 4, 5, and 6$\lambda$/D equivalent to 37, 49, 61, and 74 mas

exist, we run the simulations with 4 beam modulation amplitudes of 3, 4, 5, and 6λ/D. The trade-off is that a larger beam modulation amplitude increases the linear regime of the pyramid wavefront sensor with the drawback that the sensitivity decreases. The results for $\lambda = 3.7\mu$m are shown in Fig. 21. NCPAs with a wavefront error larger than about $0.37\mu m$ rms cannot be tolerated because the requirements are no longer met. To achieve the goal requirements of R-MET-111, NCPAs wavefront errors should be below $0.28\mu m$ rms. It is interesting to note that the modulation amplitudes do not systemically increase with increasing NCPAs.

Note that our method of looking at the Strehl ratios only may yield overly optimistic results. Even the very small drops in Strehl seen for the best seeing condition and 500 Hz loop frequency around 0.2 $\mu$m NCPA in Fig. 21 may significantly impact the high-contrast performance.

## 7 Simulation of representative METIS observations

METIS is a multi-purpose instrument. One of its main science goals is the detection and characterization of exoplanets. Here, we simulate representative observations of a planet-host star for high-contrast imaging.

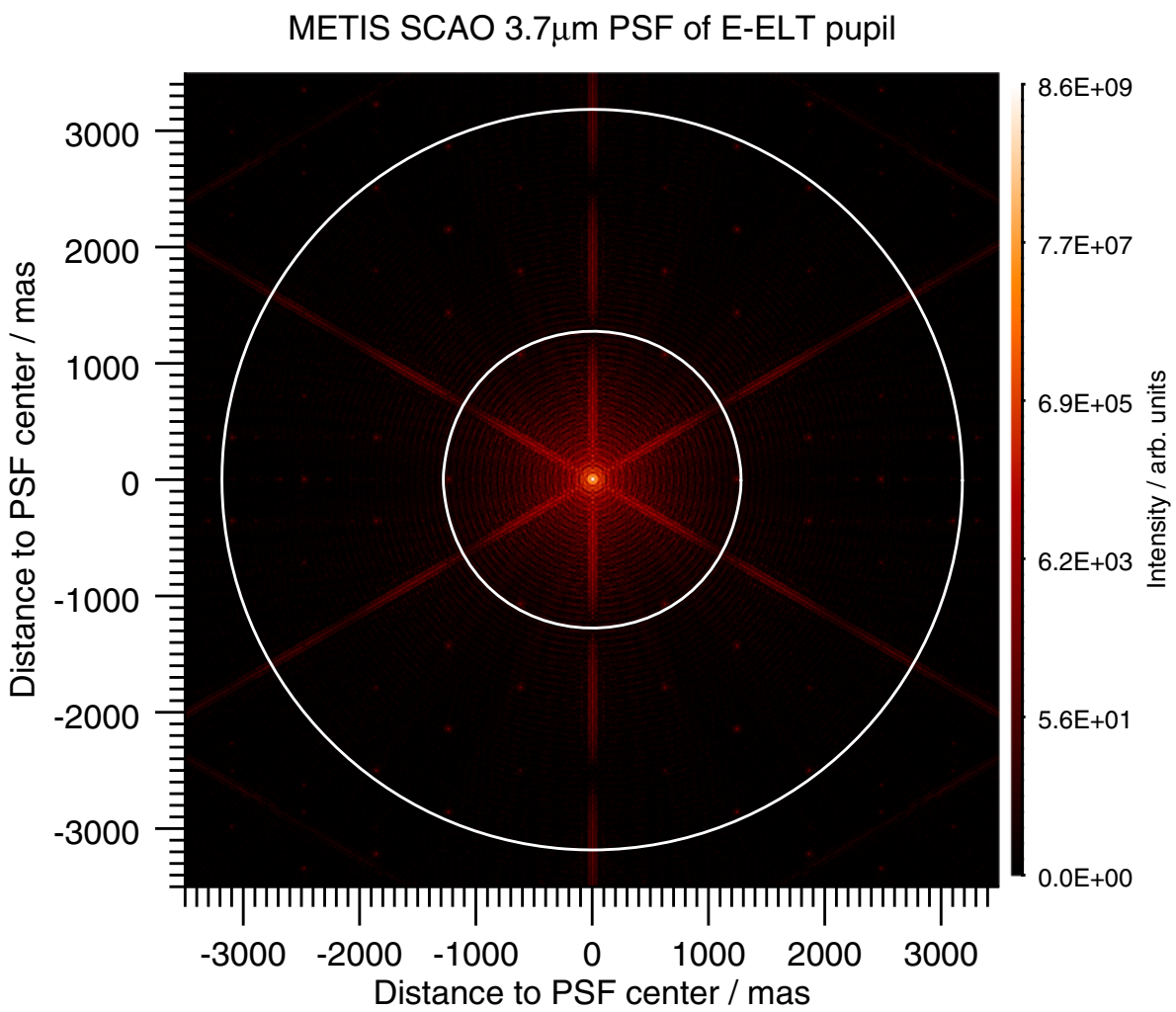

**Fig. 22** The $\lambda = 3.7\,\mu$m point spread function (PSF) of the circular ELT pupil with overlaid 60 cm spiders and $d_{seg} = 1.45$ m sized segments as shown in Fig. 8b. The angular distance of some of the expected secondary peak locations due to the ELT's primary mirror segmentation [55] is indicated with the slightly larger plotted white circles. For the inner circle the radial distance of the secondary peaks is at $r_{inner} = 4\lambda/(\sqrt{3} \cdot d_{seg})$ = 1219 mas and for the outer circle at $r_{outer} = 6\lambda/d_{seg} \cdot (1/\sqrt{6} + 1/\sqrt{3}) = 3121$ mas. To visualise the high contrast, the image has been stretched with a hyperbolic arcsine function.

### 7.1 The 51 Eridani test case

The exoplanet 51 Eridani b [53, 54] orbits the nearby star 51 Eridani with an angular separation of $\approx 0.5$ arcsec. 51 Eridani b is one of very few exoplanets with a well sampled atmospheric spectrum and therefore a good benchmark target for METIS to verify the METIS high-contrast performance using the bright, $K = 4.54$ mag host star as reference for the wavefront sensor. For the 60 min long simulations, separated in 60 yao simulations over 1 min, we used a fixed seeing of 0.64 arcsec, and a starting zenith angle of 23.325 degree. For this long simulation, instead of optimizing the control loop with respect to the vibration spectrum, we included amplitude reduced (25%) vibrations on M2, and used the yao internal stack-array deformable mirror model. The AO/DM loop gain was set to 0.1. Every second the actual residual wavefront produced by the PYR74 SCAO system was saved. We further recorded the yao reported long exposure Strehl ratio after each 1 min run. Note that the M2 vibrations repeat every 5 min.

### 7.2 Strehl ratio and PSF in direct imaging

To better understand the point spread functions (PSF) delivered by the SCAO system, we first have a look at the structure of the ELT PSF in Fig. 22. The segmented primary mirror of the ELT together with the spiders that hold the secondary mirror determine the basic PSF that the telescope can deliver. Any high-contrast imaging device has to take that into account. For our simulations, this is reflected in the used pupil mask as outlined in Section 5.4.

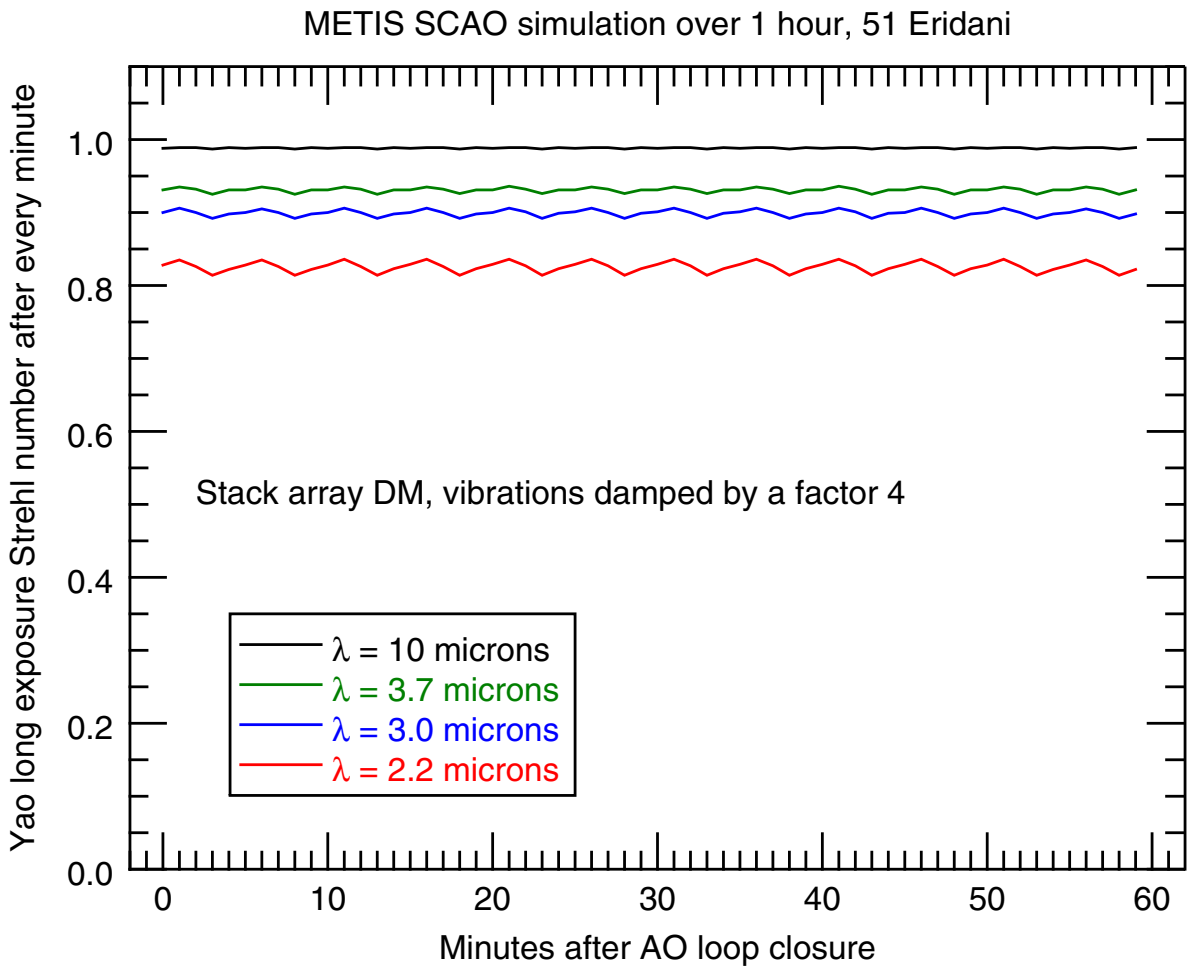

**Fig. 23** SCAO performance for the 51 Eridani test case. For this simulation we used a reduced wind induced vibration spectrum rather than optimizing the AO loop control parameters. Note that the M2 wind profile repeats every 5 minutes. The long exposure Strehl numbers are evaluated every minute, the lines are only shown to guide the eye

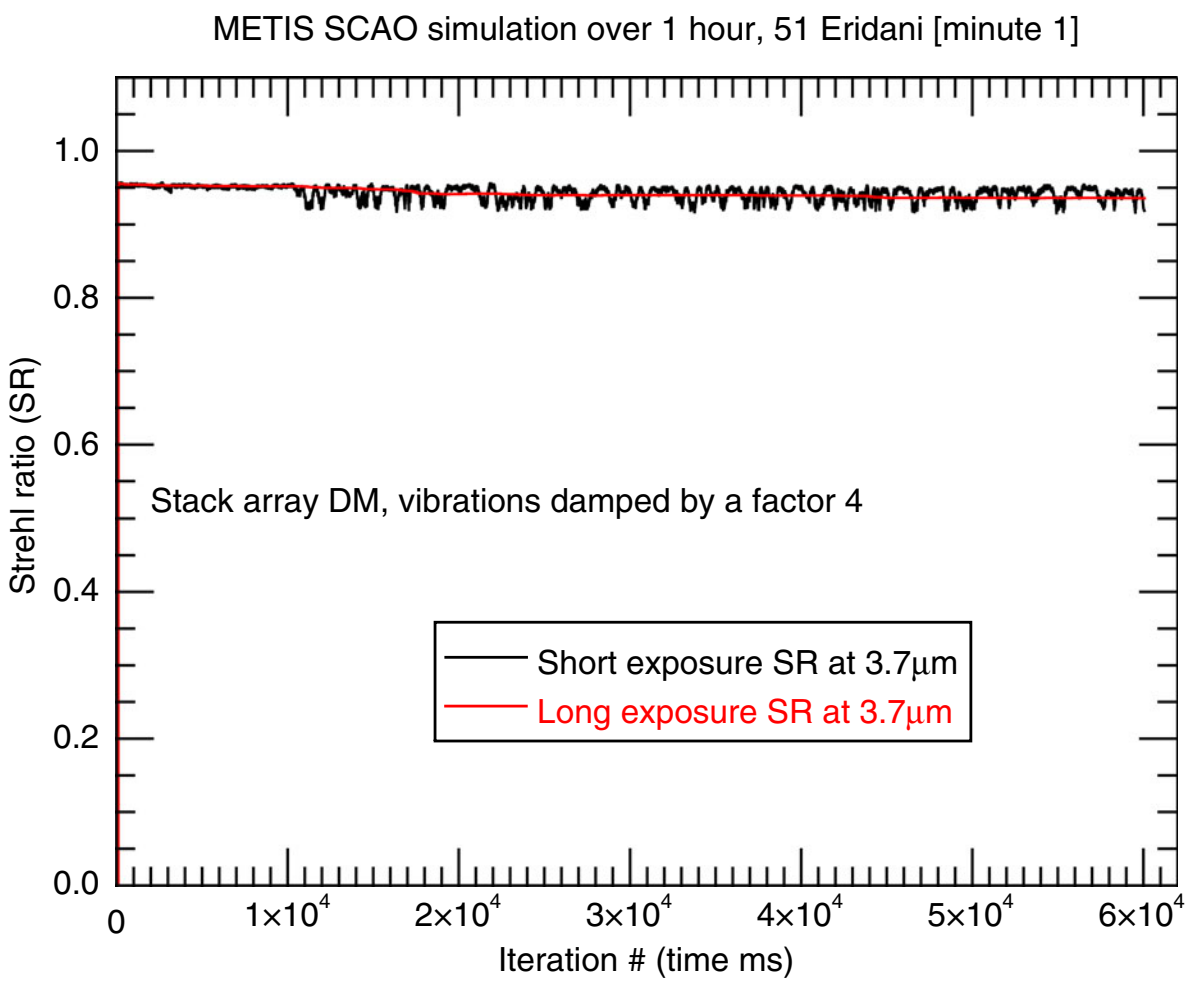

**Fig. 24** Same as Fig. 23. Short and long exposure Strehl numbers at 3.7 $\mu$m evaluated every millisecond over the first minute for the 51 Eridani test case

Figure 23 shows the long exposure Strehl ratio evaluated every minute at four different wavelengths. We note that the performance is lower in comparison to our results without M2 vibrations, which can be attributed to the higher residual image motion. The periodic structure clearly visible for the shorter wavelengths can be assigned to the M2 vibration profile, that repeats every 5 minutes. Figure 24 zooms into the first minute of the 1 hour simulation. It shows the instantaneous Strehl ratio at 3.7 $\mu$m recorded for each 1 ms iteration in yao. Over the first 10 s the M2 vibrations are very low reflected in the low variation of the instantaneous SR.

The two exemplary 51 Eri point spread functions (PSF) at $\lambda = 3.7\mu$m shown in Fig. 25 visualise that the Strehl ratio criterium is not sufficient to quantify the contrast inside the best corrected central area of the PSF. This area, defined by the actuator density of the deformable mirror, has a square shape with a side length of 74 $\lambda/d_{tel}$ = 1526 mas for the $d_{tel} = 37$ m ELT.

Within a 60 s simulation, instantaneous PSFs can vary with respect to Strehl ratio as shown in Fig. 24. Rather small variations of Strehl numbers can change the contrast at certain locations of the PSF by a factor of $\approx 10$. The maximum contrast in the PSFs of Fig. 25 is of the order $10^{-6}$ within the control radius of the deformable mirror, while the contrast between the central peak and the secondary peaks overlapping with the telescope spider structure is of the order $10^{-3}$. Looking at the instantaneous Strehl ratio variation in Fig. 24, the corresponding variation of the PSF structure during a 1 min sequence, allows for a more quantitative analysis of the high-contrast imaging performance of the METIS instrument. That means, that the SCAO delivered PSF has to be further processed using the METIS coronagraph simulation tool [56].

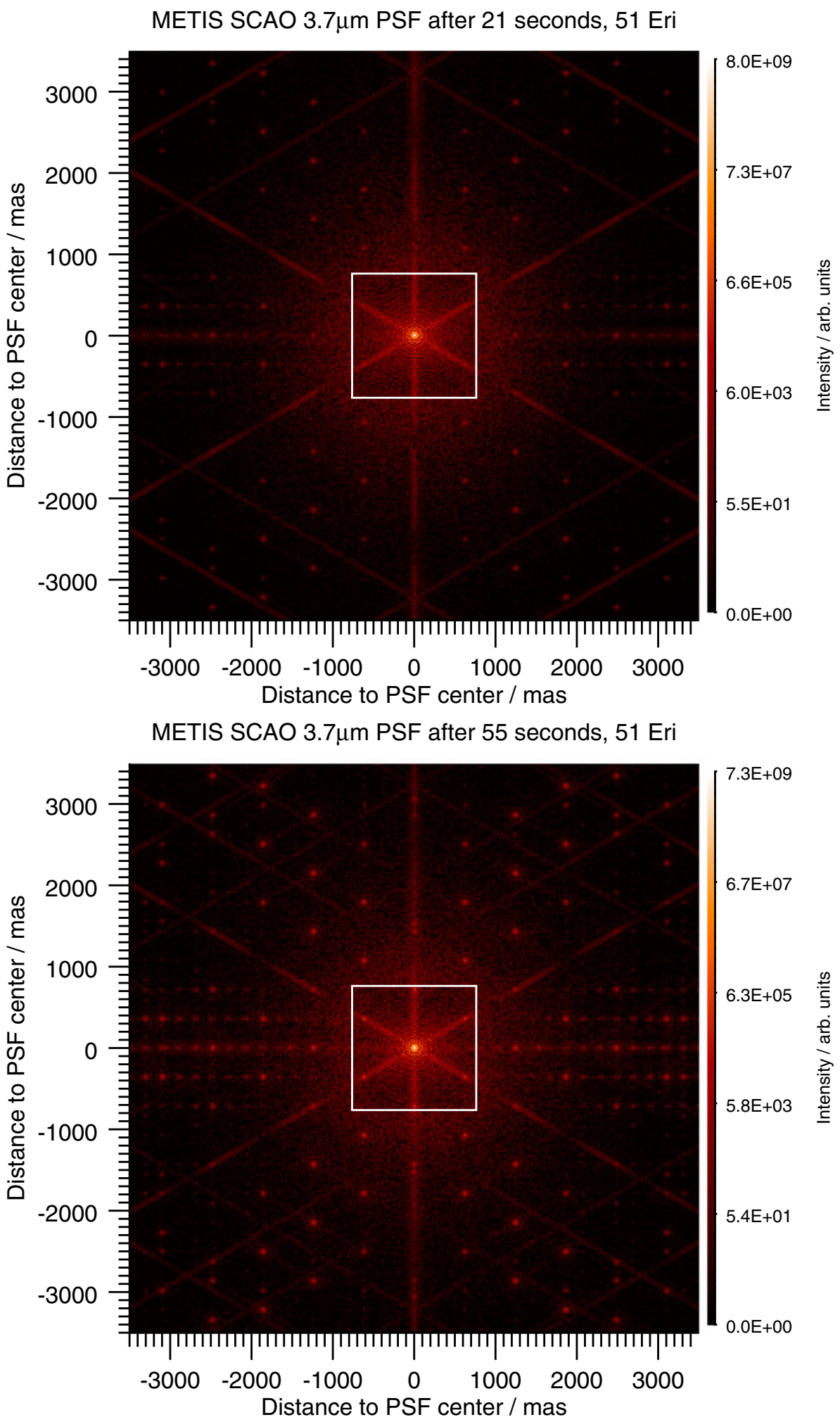

**Fig. 25** Typical SCAO point spread functions (PSF) of the PYR74 system at a wavelength of 3.7$\mu$m for the 51 Eri test case. The 1526 mas sized square box (white) shows the "control radius" of the deformable mirror. Top: PSF recorded 21s after starting the SCAO system with an instantaneous Strehl ratio of about 0.95. Bottom: PSF recorded after 55 s with a slightly worse SR. To visualise the high contrast in these images they have been stretched with a hyperbolic arcsine function

### 7.3 The structure of the METIS coronagraphic PSF

Two types of coronagraphic modes will be available in METIS, based either on a focal-plane vortex phase mask [57, 58], or an apodizing pupil-plane phase mask [59]. Here, we illustrate the capability of the ring-apodized vortex coronagraph [60], one of the baseline observing modes of METIS [56, 61], to reject light from the on-axis light during the observations of the 51 Eri exoplanetary system.

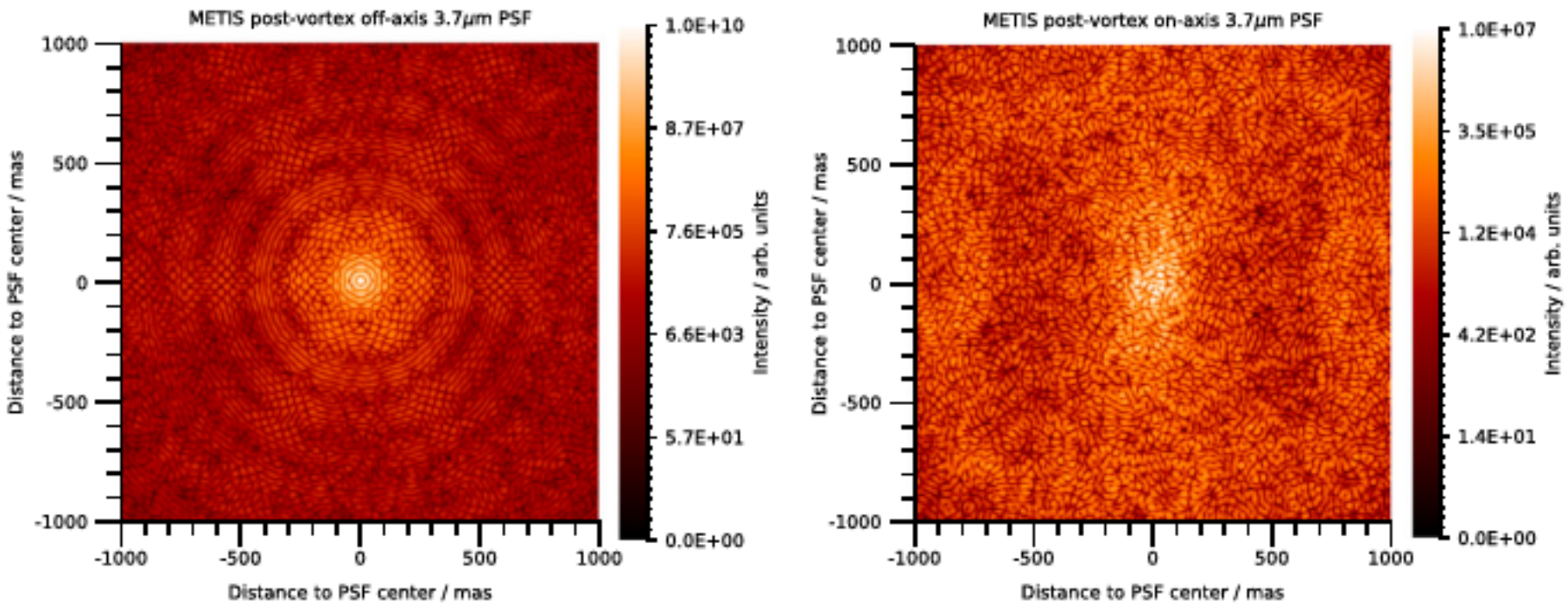

**Fig. 26** METIS post-vortex off-axis (left) and post-vortex on-axis (right) short-exposure point spread functions (PSF) of the PYR74 system at a wavelength of 3.7$\mu$m for the 51 Eri test case. To visualise the high contrast in these images they have been stretched with a hyperbolic arcsine function. See text for details

In Fig. 26, we present the short-exposure on-axis and off-axis response of the METIS ring-apodized vortex coronagraph (RAVC) to a point-like source, for a representative phase screen extracted from the SCAO simulations described in Section 7.2. Looking at short-exposure PSFs means that we take out effects of residual image jitter because we know that our simulations are not yet optimized to deal sensibly with vibrations. This helps to investigate the potential of the vortex coronagraph as we know that this device is very sensitive to image jitter.

The dimming of the bright on-axis emission produced by the coronagraph puts more clearly in evidence the structure of the SCAO residual phase screens, including the square control region associated to the DM, and the preferential direction for aberrations due to the wind (vertical in this case). Figure 27 presents the radial profiles associated with the on-axis and off-axis PSFs. It illustrates that the stellar light can be rejected by a factor ranging from 100 to 10000 in the inner 100 mas around the star, which leads to raw contrasts of the order of $10^{-4}$ at a few resolution elements from the star. This is the region where the gain in sensitivity will be the largest with the vortex coronagraph.

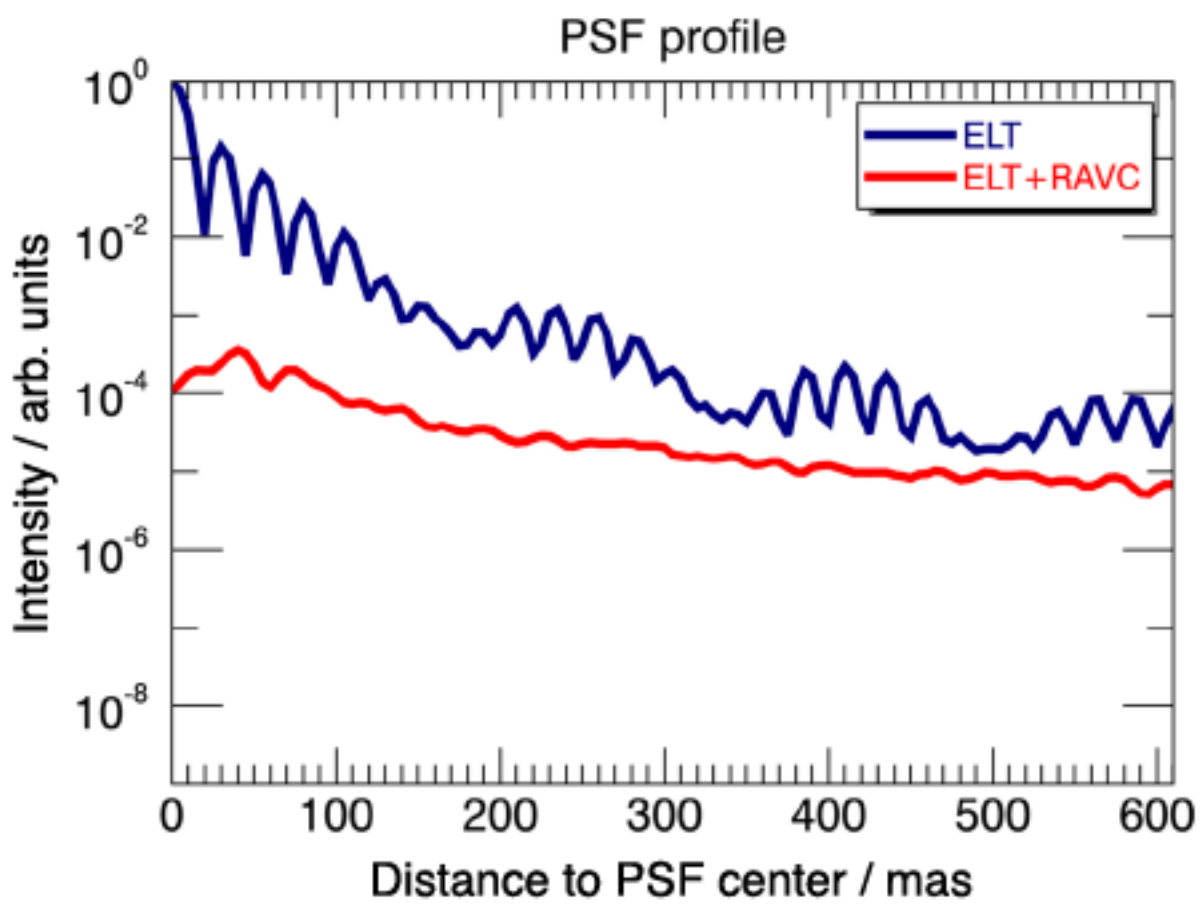

**Fig. 27** Radial profiles of the $\lambda$=3.7 $\mu$m off-axis (blue) and on-axis (red) PSFs for the ring-apodized vortex coronagraph (RAVC) on METIS using a representative SCAO residual phase screen

A more comprehensive description of the high-contrast imaging performance of METIS, including the effect of angular differential imaging on the achievable contrast after post-processing, is outside the scope of this paper, and will be presented in a forthcoming study [62]. This study will include end-to-end simulations of the two coronagraphic modes, and will also highlight the gain of the pyramid wavefront sensor in terms of achievable contrast, compared to the Shack-Hartmann wavefront sensor.

## 8 Conclusions

The goal of the work presented in this paper was to find the most suitable natural guide star wavefront sensor for the METIS instrument to be installed at the ELT. Analyzing the sky and sample coverage over the visible and near-infrared spectral range, we find that using the near-infrared band is advantageous for METIS. Whether this spectral band will cover the 1.1–2.4 $\mu$m or 1.4–2.4 $\mu$m range depends on the final optical design of the instrument.

Building on the results of previous studies [8, 14] we performed detailed adaptive optics simulations for three wavefront sensor types, two Shack-Hartmann wavefront sensors and one pyramid wavefront sensor. Among these, the pyramid wavefront sensor with 74 × 74 sub-apertures shows the best overall performance. This selection offers the advantage of using an existing, high-speed and low noise near-infrared detector with a sufficient number of pixels. Analyzing the performance of the PYR74 system under the presence of non-common path aberrations, we find that the PYR74 system provides the required performance for NCPAs up to 0.37$\mu$m rms.

For wind induced vibrations on the ELT secondary mirror, our simulations can be considered only very preliminary since an adequate adaptive vibration compensation was not available in our simulation tool.

Our 1 minute simulations show that the PYR74 system can properly reconstruct wavefronts on the fragmented ELT pupil as no piston differences buildup between the pupil fragments. The delivered AO performance stability over long periods as well as the absolute instantaneous AO performance are important requirements for the high-contrast imager of METIS. Our first results show that the simulated PYR74 system is a good baseline for METIS.

The next steps in order to create the complete error budget for the METIS SCAO system, we will investigate the AO performance including atmospheric dispersion, pupil shifts and mis-registration between actuators and sub-apertures.

The latter point, the WFS to ELT M4 DM actuator grid registration precision is an important factor of the SCAO error budget. In a preliminary study/analysis we have found that lateral pupil shifts as large as 15 cm with respect to the ELT M1 result in an unstable AO loop. For pupil rotations the limit is 0.5 degrees. Our preliminary requirement for the SCAO pupil registration precision is 5–10 cm (the typical rule of thumb is to keep the pupil registered within 1/10 the size of a sub-aperture). To maintain this precision we will implement a pupil lateral motion tracking algorithm as outlined in [63]. The large number of sub-apertures corresponding to their small size of 50 cm also limits the errors when using numerical pupil rotation.

In consideration that METIS will likely operate without a laser guide star system, which has a strong impact on sky coverage, we will balance again the number of sub-apertures (sky coverage) with the METIS science cases. We furthermore want to implement adaptive vibration compensation in our simulation tool and evaluate the PYR74 performance using a modal control scheme.

In view of a timely hardware implementation of the PYR74 system, two of the three most challenging components are already commercially available, the wavefront sensor detector and the AO real-time computer. The required cryogenic modulator needs to be build. We launched a study aimed at building such a device.

**Acknowledgements** Open access funding provided by Max Planck Society. The authors want to thank François Rigaut and Marcos van Dam for their help, support, and continuous update of the yao simulation tool. Special thanks to the reviewer for the very valuable comments and suggestions to improve the quality of this article.

The research leading to these results was partly funded by the European Research Council under the European Union's Seventh Framework Programme (ERC Grant Agreement n.337569) and by the French Community of Belgium through an ARC grant for Concerted Research Actions. OA acknowledges funding from the F.R.S.-FNRS.

## Affiliations

**Stefan Hippler**[1] · **Markus Feldt**[1] · **Thomas Bertram**[1] · **Wolfgang Brandner**[1] · **Faustine Cantalloube**[1] · **Brunella Carlomagno**[2] · **Olivier Absil**[2] · **Andreas Obereder**[3] · **Iuliia Shatokhina**[4] · **Remko Stuik**[5]

✉ Stefan Hippler
hippler@mpia.de

Markus Feldt
feldt@mpia.de

Thomas Bertram
bertram@mpia.de

Wolfgang Brandner
brandner@mpia.de

Faustine Cantalloube
cantalloube@mpia.de

Brunella Carlomagno
brunella.carlomagno@student.ulg.ac.be

Olivier Absil
olivier.absil@uliege.be

Andreas Obereder
andreas.obereder@mathconsult.co.at

Iuliia Shatokhina
iuliia.shatokhina@indmath.uni-linz.ac.at

Remko Stuik
stuik@strw.leidenuniv.nl

1 Max-Planck-Institut für Astronomie, Königstuhl 17, 69117 Heidelberg, Germany

2 Space sciences, Technologies, and Astrophysics Research (STAR) Institute, Université de Liège, 19C Allée du Six Août, B-4000 Sart Tilman, Liège, Belgium

3 MathConsult GmbH, Altenbergerstraße 69, A-4040 Linz, Austria

4 Industrial Mathematics Institute, Johannes Kepler University Linz, Altenbergerstraße 69, A-4040 Linz, Austria

5 Leiden Observatory, Leiden University, P.O. Box 9513, 2300 RA Leiden, The Netherlands